\documentclass[11pt,draftclsnofoot,peerreviewca,letterpaper,onecolumn]{IEEEtran}

\usepackage{cite}
\usepackage{graphicx,color,epsfig,rotating}
\usepackage{amsfonts,amsmath,amssymb,bbm}
\usepackage{algorithm,algorithmic}
\usepackage{subfigure}
\long\def\comment#1{}

\newfont{\bbb}{msbm10 scaled 700}

\newfont{\bb}{msbm10 scaled 1100}
\newcommand{\CC}{\mbox{\bb C}}

\newcommand{\RR}{\mbox{\bb R}}

\newcommand{\ZZ}{\mbox{\bb Z}}

\newcommand{\EE}{\mbox{\bb E}}

\newcommand{\av}{{\bf a}}

\newcommand{\ev}{{\bf e}}

\newcommand{\hv}{{\bf h}}

\newcommand{\nv}{{\bf n}}

\newcommand{\rv}{{\bf r}}

\newcommand{\vv}{{\bf v}}

\newcommand{\yv}{{\bf y}}
\newcommand{\zv}{{\bf z}}
\newcommand{\zerov}{{\bf 0}}
\newcommand{\onev}{{\bf 1}}

\newcommand{\Am}{{\bf A}}
\newcommand{\Bm}{{\bf B}}

\newcommand{\Gm}{{\bf G}}
\newcommand{\Hm}{{\bf H}}
\newcommand{\Id}{{\bf I}}

\newcommand{\Lm}{{\bf L}}
\newcommand{\Mm}{{\bf M}}

\newcommand{\Rm}{{\bf R}}

\newcommand{\Um}{{\bf U}}
\newcommand{\Vm}{{\bf V}}

\newcommand{\Xm}{{\bf X}}
\newcommand{\Ym}{{\bf Y}}
\newcommand{\Zm}{{\bf Z}}

\newcommand{\Ac}{{\cal A}}

\newcommand{\Cc}{{\cal C}}
\newcommand{\Dc}{{\cal D}}
\newcommand{\Ec}{{\cal E}}

\newcommand{\Gc}{{\cal G}}

\newcommand{\Ic}{{\cal I}}

\newcommand{\Nc}{{\cal N}}

\newcommand{\Pc}{{\cal P}}

\newcommand{\Vc}{{\cal V}}

\newcommand{\Xc}{{\cal X}}

\newcommand{\Sigmam}{\hbox{\boldmath$\Sigma$}}
\newcommand{\Phim}{\hbox{\boldmath$\Phi$}}

\newcommand{\Psim}{\hbox{\boldmath$\Psi$}}

\newcommand{\Xim}{\hbox{\boldmath$\Xi$}}

\newcommand{\diag}{{\hbox{diag}}}
\renewcommand{\det}{{\hbox{det}}}
\newcommand{\trace}{{\hbox{tr}}}

\renewcommand{\arg}{{\hbox{arg}}}

\newcommand{\SNR}{{\sf SNR}}

\newcommand{\eqdef}{\stackrel{\Delta}{=}}

\newcommand{\herm}{{\sf H}}

\newcommand{\Ltx}{L_{\rm D}}
\newcommand{\Ltr}{L_{\rm P}}

\newtheorem{example}{Example}%[section]
\newtheorem{theorem}{Theorem}%[section]
\newtheorem{lemma}{Lemma}%[section]
\begin{document}

\title{Achieving ``Massive MIMO'' Spectral Efficiency with a
Not-so-Large Number of Antennas}

\author{Hoon Huh,~\IEEEmembership{Student Member,~IEEE}, \\
Giuseppe Caire,~\IEEEmembership{Fellow,~IEEE}, \\
Haralabos C. Papadopoulos,~\IEEEmembership{Member,~IEEE}, \\
and Sean A. Ramprashad,~\IEEEmembership{Senior Member,~IEEE}
\thanks{H. Huh and G. Caire are with the Department of Electrical Engineering,
University of Southern California, Los Angeles, CA 90089, USA. (e-mail: \{hhuh, caire\}@usc.edu)}
\thanks{H. C. Papadopoulos and S. A. Ramprashad are with DOCOMO USA Labs, Palo Alto, CA 94304,
USA. (e-mail: \{hpapadopoulos, ramprashad\}@docomolabs-usa.com)}
}

\maketitle
\date{\today}

\vspace{5cm}


\newpage
\thispagestyle{empty}

\begin{abstract}
Time-Division Duplexing (TDD) allows to estimate the downlink channels for
an arbitrarily large number of base station antennas from a finite number of orthogonal
pilot signals in the uplink, by exploiting channel reciprocity.
Therefore, while the number of users per cell served in any time-frequency channel
coherence block is necessarily limited by the number of pilot sequence dimensions available,
the number of base station antennas can be made as large as desired.
Based on this observation, a recently proposed
very simple ``Massive MIMO'' scheme was shown to achieve
unprecedented spectral efficiency in realistic conditions of user spatial distribution,
distance-dependent pathloss and channel coherence time and bandwidth.
%, using a very simple beamforming
%scheme.

The main focus and contribution of this paper is a novel network-MIMO TDD architecture that achieves
spectral efficiencies comparable with ``Massive MIMO'', with one order of magnitude fewer antennas
per active user per cell.
The proposed architecture is based on a family of network-MIMO schemes defined by small clusters of cooperating base stations,
zero-forcing multiuser MIMO precoding with suitable inter-cluster interference constraints,
uplink pilot signals reuse across cells, and frequency reuse.
The key idea consists of partitioning the users population into geographically determined ``bins'',
such that all users in the same bin are statistically equivalent, and use the optimal network-MIMO architecture in the family
for each bin. A scheduler takes care of serving the different bins on the time-frequency slots,
in order to maximize a desired network utility function that captures some desired notion of fairness.
This results in a mixed-mode network-MIMO architecture, where different schemes, each of which is optimized for the served user bin,
are multiplexed in time-frequency.

In order to carry out the performance analysis and the optimization of the proposed
architecture in a clean and computationally efficient way, we consider the large-system regime where the number of users, the number of antennas,
and the channel coherence block length go to infinity with fixed ratios.
%In this regime, we show that the simple beamforming scheme previously
%advocated performs very poorly.
The performance predicted by the large-system asymptotic analysis matches
very well the finite-dimensional simulations. Overall, the system spectral efficiency obtained by the
proposed architecture is similar to that achieved by ``Massive MIMO'', with a 10-fold reduction in the number
of antennas at the base stations (roughly, from 500 to 50 antennas).
\end{abstract}

\newpage
\setcounter{page}{1}

%%%%%%%%%%%%%%%%%%%%%%%%%%%%%%%%%%%%%%%%%%%%%%%%%%%%%%
\section{Introduction}

\IEEEPARstart{M}{ultiuser} MIMO (MU-MIMO) technology is being intensively
studied for the next generation wireless cellular systems (e.g., LTE-Advanced \cite{LTE-Advanced-TR10}).
Schemes where antennas of different Base Stations (BSs) are jointly processed by centralized BS controllers
are usually referred to as ``network-MIMO'' architectures (e.g., \cite{Boudreau-CommMag09, Dahrouj-Yu-TWC10,
Huh-Caire-TSP10,Foschini-IPC06, Jing-ISIT07,Boccardi-Huang-PIMRC07, Caire-Docomo-Allerton08}).
It is well-known that the improvement obtained from transmit antenna joint processing
is limited by a ``dimensionality'' bottleneck  \cite{Ramprashad-Caire-PIMRC09,
Ramprashad-Caire-Asilomar09, Huh-Tulino-Caire-TITsubmit}.
In particular, the high-SNR capacity of a single-user MIMO system with $N_t$ transmit antennas, $N_r$ receiving antennas,
and fading coherence block length $T$ complex dimensions,\footnote{The fading coherence block
length $T$, measured in signal complex ``dimensions'' in the time-frequency domain
is proportional to the product $W_c T_c$, where $T_c$ (s) denotes the channel coherence interval, and $W_c$ (Hz) denotes the channel coherence
bandwidth \cite{Proakis-Book00}.}
scales as $C(\SNR) = \min\{N_t, N_r, T/2\} \log \SNR + O(1)$ \cite{Marzetta-Hochwald-TIT99, Zheng-Tse-TIT02}.
Therefore, even by pooling all base stations into a single distributed
macro-transmitter with $N_t \gg 1$ antennas and all user terminals into a single distributed macro-receiver with $N_r \gg 1$ antennas,
the system {\em degrees of freedom}\footnote{The system Degrees of Freedom (DoFs) are defined as
the pre-log factor of the system capacity $C(\SNR)$, i.e.,
${\rm DoFs} \; = \; \lim_{\SNR \rightarrow \infty} \frac{C(\SNR)}{\log \SNR}$, and quantify the number of ``equivalent'' parallel single-user Gaussian channels,
in a first-order approximation with respect to $\log \SNR$.}
are eventually limited by the fading coherence block length $T$.
While this dimensionality bottleneck is an inherent fact, emerging from the high-SNR behavior of the capacity
of MIMO block-fading channels,  \cite{Zheng-Tse-TIT02} (see also  \cite{Hassibi-Hochwald-TIT03})
the {\em same behavior} also characterizes the capacity scaling of MU-MIMO systems based on
explicit training for channel estimation, and  can be interpreted as the effect of the ``overhead'' incurred by
pilot signals \cite{Caire-Jindal-Kobayashi-Ravindran-TIT10}.

For frequency-division duplex (FDD) systems, the training overhead required
to collect channel state information at the transmitters (CSIT) grows linearly with
the number of cooperating transmit antennas.
%Since the number of training dimensions that can be allotted to obtain non-outdated CSIT is
%inherently limited by the channel coherence block length,
Such overhead restricts the MU-MIMO benefits that can be harvested with a large number of transmit antennas, as shown in
\cite{Ramprashad-Caire-Asilomar09} using system-level simulation and in \cite{Huh-Tulino-Caire-TITsubmit} using closed-form
analysis based on the limiting distribution of certain large random matrices.

For Time Division Duplexing (TDD) systems, exploiting {\em channel reciprocity} \cite{Marzetta-Asilomar06,Marzetta-TWC10},
%for which
%the uplink and downlink channels between a pair of antennas are strongly
%correlated within the same channel coherence block,
the CSIT can be obtained from the uplink training.
In this case, the pilot signal overhead scales linearly with the number of {\em active users} per cell,
but it is independent of the number of cooperating antennas  at the BSs.
As a result, for a fixed number of users scheduled for transmission,
the TDD system performance can be significantly improved by increasing the number of BS antennas.

Following this idea, Marzetta \cite{Marzetta-TWC10} has shown that simple {\em Linear Single-User BeamForming}
(LSUBF) and random user scheduling, without any inter-cell cooperation, yields unprecedented
spectral efficiency in TDD cellular systems, provided that a {\em sufficiently} large number of transmit antennas
per active user are employed at each BS. This scheme, nicknamed hereafter ``Massive MIMO'',
was analyzed in the limit of infinite number of BS antennas per user per cell. In this regime, the effects of Gaussian noise and uncorrelated inter-cell interference
disappear, and that the only remaining impairment is the inter-cell interference
due to {\em pilot contamination} \cite{Jose-Ashikhmin-Marzetta-Vishwanath-TWC11},
i.e., to the correlated interference from other cells due to users re-using the same pilot signal (see Section \ref{sec:training}).
%%%%%%%%%%%%%%%%%%%%%%%%%%%%%%%%%%%%%%%%%%%%%%%%%%%%%%%%%%%%%%%%%%%%%%%%%%%%%%%
%comes from reusing pilot signals in other cells, which is inherently inevitable in a cellular system with a finite number of dimensions per
%coherence block. Consider a TDD system serving $S$ users per cell. In order to obtain
%the corresponding CSIT by exploiting the TDD reciprocity, $S$ orthogonal pilot signals must be sent in every channel coherence block
%in the uplink. This requires that at least $S$ dimensions out of the $L$ dimensions per coherence block are used by the pilot signals.
%Eventually, the same pilot signals are reused in other cells, and the CSIT is ``contaminated'' by the interference from the pilot signals reused
%in other cells. Hence,  the contaminated beamforming vectors focus some transmit power in the direction  of the unintended users
%in the other cells sharing the same pilot signals. This effect is particularly evident for users at the cell edge.
%%%%%%%%%%%%%%%%%%%%%%%%%%%%%%%%%%%%%%%%%%%%%%%%%%%%%%%%%%%%%%%%%%%%%%%%%%%%%%%%
%To mitigate the effect of pilot contamination and therefore to improve the
%cell-edge throughput, \cite{Marzetta-TWC10} considers frequency reuse with factors 3 or 7 for
%the hexagonal cell layout.
%%%%%%%%%%%%%%%%%%%%%%%%%%%%%%%%%%%%%%%%%%%%%%%%%%%%%%

In this work, we also focus on TDD systems and exploit reciprocity.
The main contribution of this paper is a novel network-MIMO architecture
that achieves spectral efficiencies comparable with ``Massive MIMO'', with a more practical number of BS antennas per active user
(one order of magnitude less  antennas for approximately the same spectral efficiency).
As in \cite{Marzetta-TWC10}, we also analyze the proposed system in the limit of a large number
of antennas. However, a different system scaling is considered,  where the number of antennas per active user per
cell is {\em finite}. This is obtained by letting the number of users per cell, the number of antennas per BS, and the channel coherence block length
go to infinity, with fixed ratios \cite{Huh-Tulino-Caire-TITsubmit}.  We find that in this regime the
LSUBF scheme advocated in \cite{Marzetta-TWC10} performs very poorly.
In contrast, we consider a family of network-MIMO schemes based on
small clusters of cooperating base stations, {\em Linear Zero-Forcing BeamForming} (LZFBF)
with suitable inter-cluster interference constraints,  uplink pilot signals reuse across cells, and frequency reuse.
The key idea consists of partitioning the users population into geographically determined ``bins'', containing  statistically equivalent users,
and optimizing the network-MIMO scheme for each individual bin.  Then, users in different bins are scheduled over the time-frequency slots, in order to maximize
an appropriately chosen network utility function reflecting some desired notion of ``fairness''.
The geographic nature of the proposed scheme yields very simple system operations, where
each time a given bin is scheduled, a subset of {\em active users} in the selected bin is chosen
at random or in a deterministic round robin fashion,  without performing any CSIT-based user selection.
This allows a fast turn-around between feedback and transmission, that can take place in the same channel coherence block.
The resulting architecture is a mixed-mode network-MIMO, where different schemes, each of which is optimized for the served user bin,
are multiplexed in time-frequency.

Using results and tools from the large-system analysis developed in  \cite{Huh-Tulino-Caire-TITsubmit, Tulino-Book04} and adapted to the present scenario,
we obtain the asymptotic achievable rate for each scheme in closed form.
The performance predicted by the large-system analysis match
very well with finite-dimensional simulations, in agreement with \cite{Huh-Tulino-Caire-TITsubmit,Huh-Caire-TITsubmit}
and with several well-known works on single-user MIMO in the large antenna regime
\cite{Moustakas-TIT03,Kumar-Caire-TIT09}.
The large-system analysis developed here is instrumental to the systematic design and optimization of the proposed system architecture,
since it allows an accurate and rapid selection of the best network-MIMO scheme for each user bin
without resorting to cumbersome and time-consuming Monte Carlo simulation. In fact, the system parameters in the considered family
of network-MIMO schemes  are strongly mutually dependent, and the system optimization without the analytical tools developed here
would just be infeasible.

%%%%%%%%%%%%%%%%%%%%%%%%%%%%%%%%%%%%%%%%%%%%%%%%%%%%%%%%%%%%%%%%%%%%%%%%%
%%%%%%%%%%%%%%%%%%%%%%%%%%%%% INSERT A PARAGRAPH ABOUT EXISTING LITERATURE %%%%%%%%%%%%%
%%%%%%%%%%%%%%%%%%%%%%%%%%%%%%%%%%%%%%%%%%%%%%%%%%%%%%%%%%%%%%%%%%%%%%%%%
We hasten to say that the ideas of dynamic clustering of cooperating BSs, and multimodal MU-MIMO downlink have appeared
in a large number of previous works (see for example \cite{forenza-gesbert-multimode,gesbert-dynamic,andrews-heath-multimode,ShiraniMehr-Caire-Neely-TCOM10,Papadopoulos-Caire-Asilomar10}.
Giving a fair account of this vast literature would be impossible within the space limits of this paper.
Nevertheless, we wish to stress here that the novel contribution of this paper is a {\em systematic approach} to multi-modal
system optimization  based  on simple closed-form expressions of the spectral efficiency of
each network-MIMO scheme in the family, and on scheduling across the schemes (or ``modes'') in order to maximize
a desired network utility function.
%%%%%%%%%%%%%%%%%%%%%%%%%%%%%%%%%%%%%%%%%%%%%%%%%%%%%%%%

The remainder of this paper is organized as follows. In Section~\ref{sec:model}, we describe
the family of proposed network-MIMO schemes.
We discuss the uplink training, MMSE channel estimation and pilot contamination effect
for TDD-based systems in Section~\ref{sec:training}.
In Section~\ref{sec:precoding}, we analyze the network-MIMO architectures under considerations
and and provide expressions for their achievable rate in the large-system limit.
Scheduling under specific fairness criteria and the corresponding system spectral efficiency is presented
in Section~\ref{sec:scheduling}. Numerical results including comparison with finite dimensional simulation
results are presented in Section~\ref{sec:results}
and concluding remarks are given in Section~\ref{sec:conclusions}.

%%%%%%%%%%%%%%%%%%%%%%%%%%%%%%%%%%%%%%%%%%%%%%%%%%%%%%%
\section{System model} \label{sec:model}

The TDD cellular architecture for high-data rate downlink proposed in this work is based on the following elements:
\begin{enumerate}
\item A family of network-MIMO schemes, defined in terms of the size and shape of clusters of cooperating BSs,
pilot reuse across clusters, frequency reuse factor, and downlink linear precoding scheme;
\item A partitioning of the user population into bins, according to their position in the cellular coverage area;
\item The determination of the optimal network-MIMO scheme in the family for each user bin,
creating an association between user bins and network-MIMO schemes;
\item Scheduling of the user bins in time-frequency in order to maximize a suitable concave and componentwise non-decreasing
network utility function of the ergodic user rates. The network utility function is chosen in order to reflect some desired notion
of fairness  (e.g., proportional fairness \cite{Mo-TNET00, Viswanath-Tse-Laroia-TIT02, Huh-Caire-TITsubmit, Bender-Viterbi-CommMag00, Parkvall-Englund-Lundevall-Torsner-CommMag06}). When a given bin is scheduled, the associated optimized network-MIMO architecture is used.
\end{enumerate}
Invoking well-known convergence results \cite{Moustakas-TIT03, Kumar-Caire-TIT09}, we use
the ``large-system analysis'' approach for multi-antenna cellular systems pioneered in
\cite{Aktas-TIT06, Hoydis-Debbah-TSPsubmit, Zakhour-Hanly-JSACsubmit, Huh-Caire-TITsubmit}. In particular, we use the results
of \cite{Huh-Tulino-Caire-TITsubmit}, which can be easily applied to our system model,  and analyze the performance of the network-MIMO schemes
in the considered family while scaling the number of users in each bin, the number of antennas per BS and the small-scale fading coherence block length
to infinity, with fixed ratios. We define a {\em system size} parameter indicated by $N$, and let all the above quantities scale
linearly with $N \rightarrow \infty$.  Specifically, we let $MN$ denote the number of BS antennas,
$LN$ denote the channel coherence  block length, and $UN$ denote the number of users per location (a bin is defined as a set of discrete locations in the cellular
coverage, see Section \ref{subsec:layout}), for given constants $M, L$ and $U$.

%%%%%%%%%%%%%%%%%%%%%%%%%%%%%%%%%
\subsection{Cellular layout and frequency reuse} \label{subsec:layout}

{\bf Base stations, cells and clusters:}
The system geometry is concisely described by using lattices on the real line $\RR$ (for 1-dimensional layouts \cite{Caire-Docomo-Allerton08, Ramprashad-Caire-Asilomar09})
or on the real plane $\RR^2$ (for 2-dimensional layouts \cite{Marzetta-TWC10}).
Consider nested lattices $\Lambda \subseteq \Lambda_{\rm bs} \subseteq \Lambda_{\rm u}$ in $\RR$ (resp., $\RR^2$).
The system {\em coverage region} is given by the Voronoi cell $\Vc$ of $\Lambda$ centered at the origin.\footnote{The Voronoi cell of a lattice point $x \in \Lambda \in \RR^n$
is the set of points $y \in \RR^n$ closer to $x$ than to any other lattice point.}
BSs are located at points $b \in \Lambda_{\rm bs} \cap \Vc$. The finer lattice $\Lambda_{\rm u}$ defines a grid of discrete
{\em user locations}, as explained later in this section.
We let $B = |\Lambda_{\rm bs} \cap \Vc|$ denote the number of BSs in the system.

\begin{example} \label{1-dim-example}
Consider the 1-dimensional layout defined by  $\Lambda = B\ZZ$ and $\Lambda_{\rm bs} = \ZZ$.
The coverage region is $\Vc = [-B/2,B/2)$ and the BS locations $b$ are given
by all integer-coordinate points in the interval $[-B/2,B/2)$. \hfill $\lozenge$
\end{example}

\begin{figure}
  \centering
  \includegraphics[width=10cm]{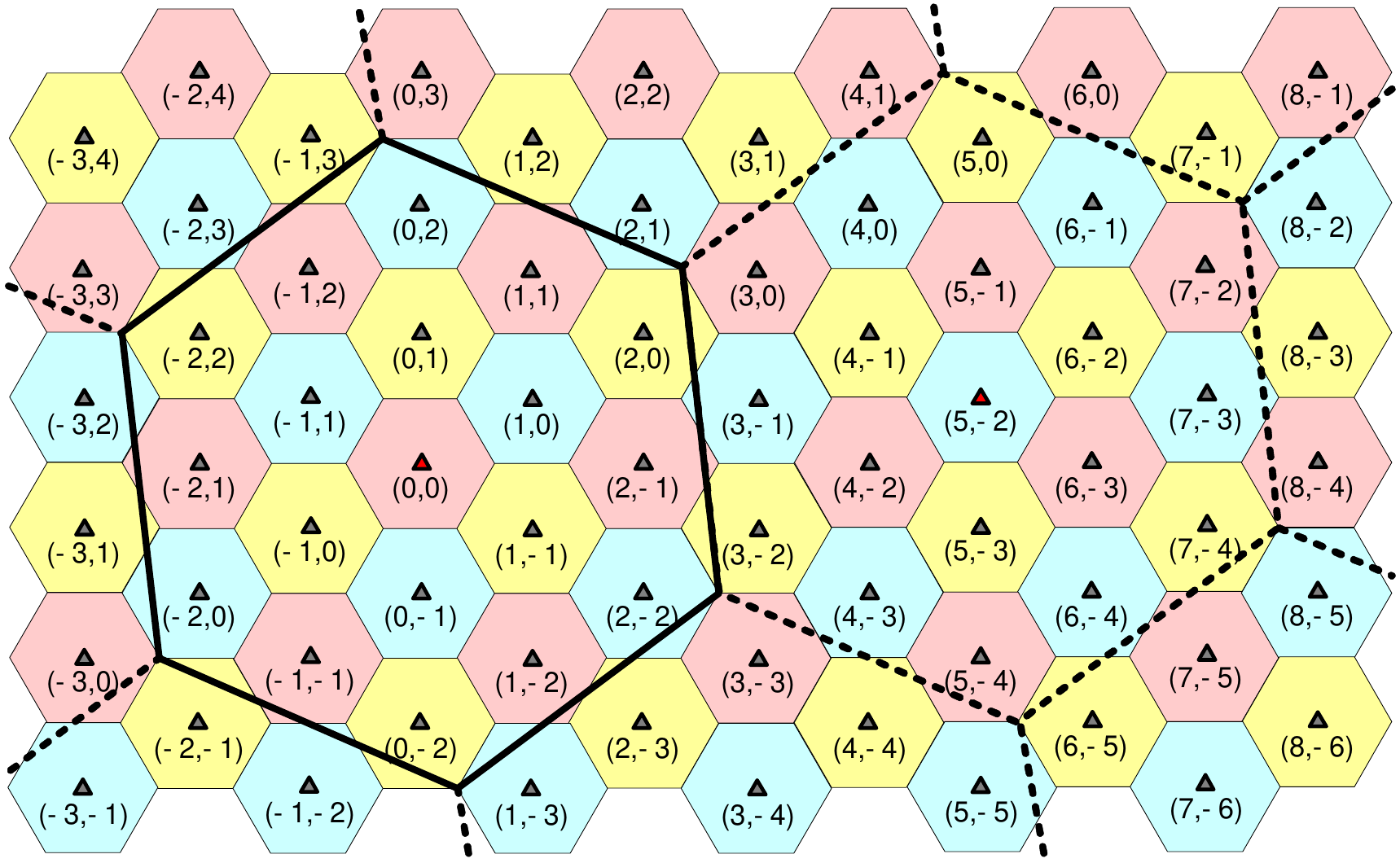}
  \caption{Two dimensional hexagonal cell layout with $B=19$. The triangle marks
  indicate the BS positions (points of $\Lambda_{\rm bs}$) and the red triangle marks
  indicate the points of $\Lambda$. The insides of the large thick-lined hexagon and the small hexagons denote
  $\Vc$ and $\Vc_b$ for $b \in \Lambda_{\rm bs}$, respectively.}
  \label{fig:2dim-lattice}
\end{figure}

\begin{example} \label{2-dim-example}
In system studies reported in the standardization of 4th generation cellular systems
\cite{LTE-Advanced-Phy-TR10, IEEE80216m-EMD-TR09} it is customary
to consider a 2-dimensional hexagonal layout formed by 19 cells, as shown in Fig.~\ref{fig:2dim-lattice}. In this case,
$\Lambda_{\rm bs} = \Am \ZZ^2$ and $\Lambda = \Am \Bm \ZZ^2$, with
$$\Am = \frac{\sqrt{3}r}{2} \begin{bmatrix} \sqrt{3} & 0\\ 1 & 2\end{bmatrix} \;\;
\mbox{and} \;\; \Bm = \begin{bmatrix} 4 & \sqrt{3}\\ -\sqrt{3} & 4\end{bmatrix},$$
where $r$ denotes the distance between the center of a small hexagon  and one of its vertices.
We have $B  = \det(\Bm) = 19$, and the distance between the closest two points in $\Lambda_{\rm bs}$ is $\sqrt{3}r$. \hfill $\lozenge$
\end{example}

For the sake of symmetry, in order to avoid ``border effects'' at the edges of the coverage region,
all distances and all spatial coordinates are defined modulo $\Lambda$.
The modulo $\Lambda$ distance between two points $u, v$ in $\RR$ (resp., $\RR^2$) is defined as
\begin{equation} \label{eq:distance}
d_{\Lambda}(u,v) = | u - v \;\; {\rm mod} \;\; \Lambda |,
\end{equation}
where $x \; {\rm mod} \; \Lambda = x - \arg\min_{\lambda \in \Lambda} |x - \lambda|$.
Cell $\Vc_b$ is defined as the Voronoi region of BS $b \in \Lambda_{\rm bs} \cap \Vc$
with respect to the modulo-$\Lambda$ distance, i.e.,
$\Vc_b = \left \{ x \in \RR : d_\Lambda(x, b) \leq d_\Lambda(x, b'),
  \;\; \forall \; b' \in \Lambda_{\rm bs} \cap \Vc \right \}$ (replace $\RR$ with $\RR^2$ for the 2-dimensional case).
The collection of cells $\{\Vc_b\}$ forms a partition of $\Vc$ into congruent regions.

\begin{figure}
  \centering
  \includegraphics[width=12cm]{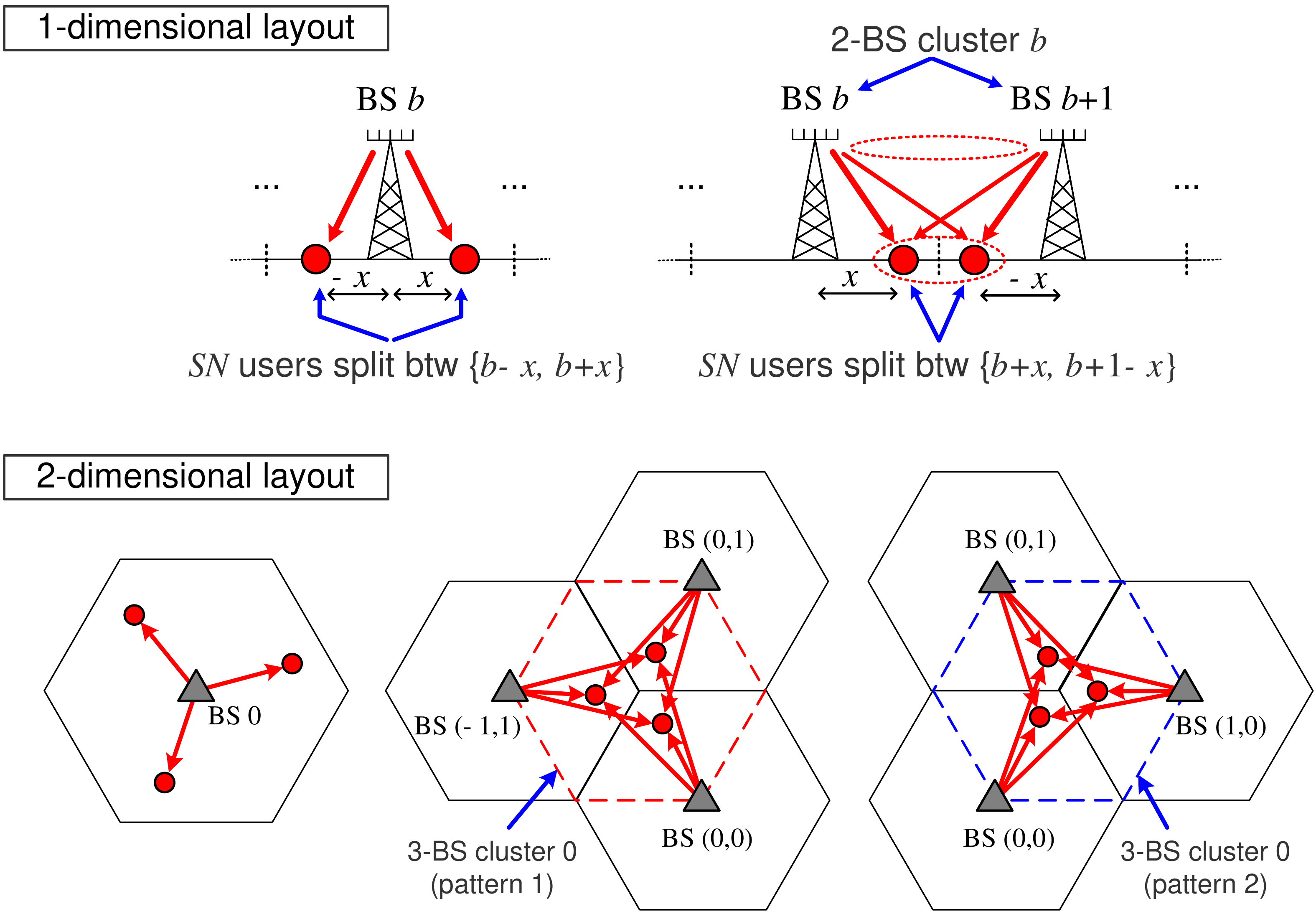}
  \caption{Cluster pattern geometry and user bins in one-dimensional and
  two dimensional layouts.}
  \label{fig:cell-cluster-bin}
\end{figure}

A ``clustering pattern'' $u(\Cc)$, defined by the set of BS locations $\Cc = \{b_0, \ldots, b_{C-1}\}$ with $b_j \in \Lambda_{\rm bs}\cap \Vc$ and
rooted at $b_0 = 0$,  is the collection of BS location sets (referred to as ``clusters'' in the following)
\begin{equation} \label{clusterc}
u(\Cc) = \{ \{ \Cc + c\} \; : \; c \in \Lambda_{\rm bs} \cap \Vc \}.
\end{equation}
We focus on systems based on single-cell processing ($C=1$), or with
joint processing over clusters of small size: $C=2$ in the 1-dimensional case,
and size $C = 3$ in the 2-dimensional case,  as shown in Fig.~\ref{fig:cell-cluster-bin}.
It turns out that larger clusters do not achieve better performance due to the large training
overhead incurred, while requiring higher complexity. Therefore, our results are not restrictive in terms of cluster size,
since they capture the best system parameters configurations.

\begin{example} \label{cluster-example}
In the 1-dimensional case of Example \ref{1-dim-example}, with $C=1$ and $C = 2$, we have
\[  u(\{0\}) = \{ \{0\}, \{1\}, \ldots,  \{B-1\}\}. \]
and
\[  u(\{0,1\}) = \{ \{0,1\}, \{1,2\}, \ldots, \{B-1,0\}\}, \]
respectively.
\hfill $\lozenge$
\end{example}

{\bf User location bins:} We assume a uniform user spatial distribution over the coverage region.
For the sake of analytical simplicity, we discretize the user distribution into a regular grid
of user locations, corresponding to the points of the lattice translate $\widetilde{\Lambda}_{\rm u} = \Lambda_{\rm u} + u_0$,
where $u_0 \neq 0$ is chosen such that $\widetilde{\Lambda}_{\rm u}$ is symmetric with respect to the origin and no points
of $\widetilde{\Lambda}_{\rm u}$ fall on the cell boundaries.

\begin{example} \label{user-locations-1dim}
In the 1-dimensional case of Example \ref{1-dim-example} we can choose $\Lambda_{\rm u} = \frac{1}{K} \ZZ$, for some even integer $K$,
and let $u_0 = \frac{1}{2K}$. Then, the points of $\widetilde{\Lambda}_{\rm u} \cap \Vc_b$ are symmetrically located
with respect to each BS of coordinate $b = 0,1,\ldots, B-1$. \hfill $\lozenge$
\end{example}

A ``user bin'' $v(\Xc)$, defined by the set of user locations $\Xc = \{x_0,x_1, \ldots, x_{m-1}\}$ with $x_i \in \widetilde{\Lambda}_{\rm u}$,
is the collection of user location sets (indicated by ``groups'' in the following)
\begin{equation} \label{binx}
v(\Xc) = \{ \{ \Xc + c\} \; : \; c \in \Lambda_{\rm bs} \cap \Vc \}.
\end{equation}
In particular, we choose $\Xc$ to be a symmetric set of points with respect to the positions of the BSs comprising
cluster $\Cc$. The reason for this symmetry is two-fold: on one hand,
a symmetric set generalizes the single location case and yet provides a set of statistically equivalent users
(same set of distances from all BSs in the cluster), thus providing a richer system optimization parameter space.
On the other hand, symmetry yields very simple closed-form expressions in the large-system analysis, by means of
\cite[Th. 3]{Huh-Tulino-Caire-TITsubmit}.

\begin{example} \label{binX-example}
In the 1-dimensional case of Example \ref{1-dim-example}, we are interested in the cases
$\Xc = \{-x, x\}$ and $\Xc = \{x,1 - x\}$, for some $x \in \widetilde{\Lambda}_{\rm u} \cap [0, 1/2]$,
as shown in Fig.~\ref{fig:cell-cluster-bin}. This yields the bins
\[ v(\{-x,x\}) = \{ \{-x,x\}, \{1-x, 1+x\}, \ldots, \{B-1-x, B-1+x\} \} \]
and
\[ v(\{x,1-x\}) = \{ \{x,1-x\}, \{1+x, 2-x\}, \ldots, \{B-1+x, B-x\} \},  \]
respectively. \hfill $\lozenge$
\end{example}

{\bf Cluster/group association and user group rate:}
The BSs forming a cluster are jointly coordinated by a ``cluster controller''
that collects all relevant channel state information
and computes the beamforming coefficients for the desired MU-MIMO precoding scheme.
For given sets $\{\Xc,\Cc\}$, the users in group $\Xc + c$
are served by the cluster $\Cc + c$, for all $c \in \Lambda_{\rm bs} \cap \Vc$ (see Fig.~\ref{fig:cell-cluster-bin}).
By construction, each BS belongs to $C$ clusters and transmits signals from all the $C$ corresponding cluster controllers.
These signals may share the same frequency band, or be defined on orthogonal subbands,
depending on the system frequency reuse factor  defined later in this section.
There are $m UN$ users in each group $\Xc+c$,  and $CMN$ jointly coordinated antennas in each cluster $\Cc+c$.
We assume $mU \geq CM$, such that the downlink DoFs are always limited by the number of antennas.\footnote{A system with
$mU < CM$ is not fully loaded, in the sense that the infrastructure would support potentially a larger number of users.}
The number of users effectively scheduled and
served on each given slot is denoted by  $S N$. We refer to these users as the ``active users'', and to the
coefficient $S \in [0,CM]$ as the ``loading factor''.
Depending on the geometry of $\Xc$ and $\Cc$ and on the type of beamforming used (see Section \ref{sec:precoding})
$S$  can be optimized for each pair $\{\Xc,\Cc\}$. We restrict to consider schemes that serve an equal number $S N/m$
of active users per location $x \in \Xc+c$.  As anticipated before, by symmetry the users in the same bin
are statistically equivalent. Therefore, without loss of generality, we may assume that a round-robin
scheduling picks all subsets of size $S N$ out of  the whole $mUN$ users in each group with the same fraction
of time. In this way,  the aggregate spectral efficiency of the group (indicated in the following a ``group spectral efficiency'')
is shared evenly among all the users in the group.

{\bf Frequency reuse:}
The frequency reuse factor of the scheme is denoted by $F$. This can also be optimized for
each given pair $\{\Xc,\Cc\}$. The system bandwidth is partitioned into $F$ subbands of equal width.
For $F = 1$, all clusters in $u(\Cc)$ transmit on the whole system bandwidth.
For $F > 1$, clusters are assigned different subbands according
to a regular reuse pattern. For the 1-dimensional layout, any integer $F$ dividing $B$ is possible.
For the 2-dimensional layout, we consider reuse factors given by
$F = i^2+ij+j^2$  for non-negative integer $i$ and $j$ \cite{Rappaport-Book02}.
For later use, we define $\Dc(f)$ as the set of clusters active on subband $f \in \{0,\ldots, F-1\}$.

\begin{figure}
  \centering
  \includegraphics[width=12cm]{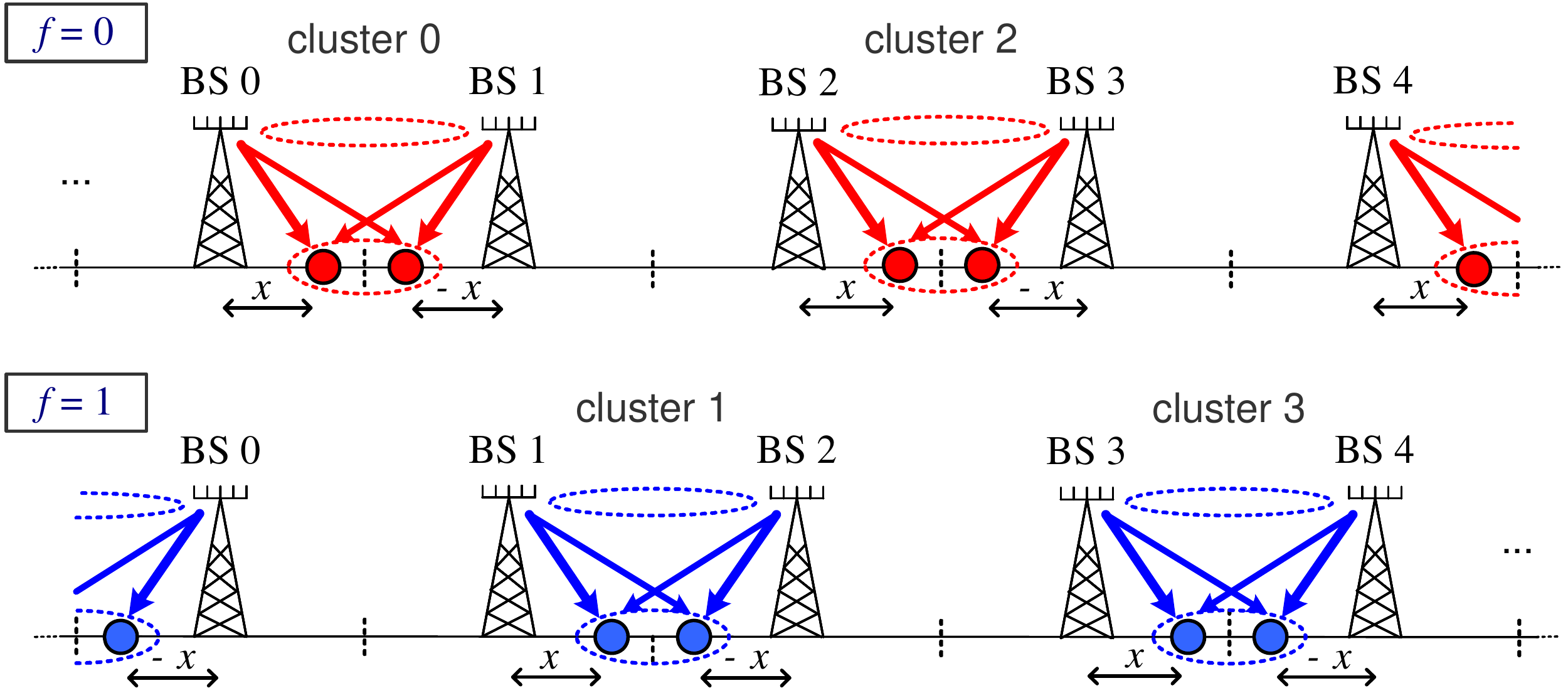}
  \caption{1-dimensional layout with $C=2$ and $F=2$.}
  \label{fig:freq-reuse}
\end{figure}

\begin{example} \label{freq-reuse-example}
Fig.~\ref{fig:freq-reuse} shows a 1-dimensional system with frequency reuse $F = 2$
for the clustering pattern of size $C=2$ defined by $\Cc=\{0,1\}$ and the user bin
defined by $\Xc = \{x, 1-x\}$. Even-numbered clusters operate on subband 0 and odd-numbered clusters operate
on subband 1. An example for the 2-dimensional hexagonal layout with
$F = 3$ and $C = 1$ is shown in Fig.~\ref{fig:2dim-lattice}, where cells with the same color operate on
the same subband. \hfill $\lozenge$
\end{example}

%%%%%%%%%%%%%%%%%%%%%%%%%%%%%%%%%%%%
\subsection{Channel statistics and received signal model} \label{subsec:channel}

The average received signal power for a user located at $x \in \Vc$ from a BS antenna
located at $b \in \Vc$ is denoted by $g(x,b)$, a polynomially decreasing function
of the distance $d_\Lambda(x,b)$. The AWGN noise power spectral
density is normalized to 1.  For fixed
For a given clustering pattern $u(\Cc)$ and user bin $v(\Xc)$,  the fading channel coefficients from the $CMN$
antennas of BS cluster $\Cc + c$ to an active user $k \in \{1,\ldots, S/m\}$ at location $x + c' : x \in \Xc$,
on frequency subband $f$, form a random vector indicated by $\underline{\hv}_{k,c',c}(f;x) \in \CC^{CMN \times 1}$, with
circularly-symmetric complex Gaussian entries, i.i.d. across the BS antennas,
the subbands and the users (independent small-scale Rayleigh fading).
In the considered network-MIMO schemes, active users are served with equal transmit power equal to $1/S$.
Hence, the total transmit power per cluster is equal to $N$. Since each BS simultaneously participates
in $C$ clusters,  also the total transmit power per BS is equal to $N$.  Since we consider the limit for $N \rightarrow \infty$,
the channel coefficients are normalized to have variance $1/N$, such that the received signal power is independent of $N$.
This provides the correct scaling of the elements of the random channel matrices in order to obtain the large-system limit results.
We let the channel vector covariance matrix be given by  $\EE\left [ \underline{\hv}_{k,c',c}(f;x) \underline{\hv}_{k,c',c}^\herm(f;x) \right ] = \frac{1}{N} \Gm_{c',c}(x)$,
where $\Gm_{c',c}(x) = \diag\left ( g(x+c',b+c) \Id_{MN} : b \in \Cc \right )$.\footnote{We use $\diag(\Mm_a : a \in \Ac)$ to indicate
a block-diagonal matrix with diagonal blocks $\Mm_a$, for some index $a$ taking values in the ordered set $\Ac$,
and $\Id_n$ to indicate the $n \times n$ identity matrix.}
Notice that $\Gm_{c',c}(x)$ is independent of the user index $k$ and on the subband index $f$, since
the channels are identically distributed across subbands and co-located users.

Under the standard block-fading assumption \cite{Proakis-Book00,Marzetta-Hochwald-TIT99, Zheng-Tse-TIT02,Marzetta-TWC10,Huh-Tulino-Caire-TITsubmit},
the channel vectors are constant on each subband for blocks of length $LN$ signal dimensions. Without loss of generality,
we assume that these coherence blocks also correspond to the scheduling slot. Each slot is partitioned into an uplink training phase,
of length $\Ltr N$ and a downlink data phase, of length $\Ltx N$.  In this section we deal with the data phase,  while the training phase is addressed in Section~\ref{sec:training}.
For the sake of notation simplicity, the slot ``time'' index is omitted: since we care about ergodic (average) rates, only the per-block marginal
channel statistics matter. The data-bearing signal transmitted by cluster $\Cc + c$ on subband $f$  is denoted by
\begin{equation} \label{tx-signal}
\Xm_c(f) = \Um_c(f) \Vm^\herm_c(f)
\end{equation}
where the matrix $\Um_c(f) \in \CC^{\Ltx N \times S N}$ contains the codeword (information-bearing) symbols arranged by columns.
We assume that users' codebooks are drawn from an i.i.d. Gaussian random coding ensemble with symbols $\sim \Cc\Nc(0,1/S)$.
Achievable rates shall be obtained via the familiar {\em random coding argument} \cite{Cover-Thomas-Book05}
with respect to this input distribution.
The matrix $\Vm_c(f)  \in \CC^{CMN \times SN}$ contains the beamforming vectors arranged
by columns, normalized to have unit norm. It is immediate to verify that, indeed, the average transmit power of any cluster $\Cc+c$,
active on frequency $f$, is given by
%\begin{eqnarray} \label{power-norm}
%\frac{1}{\Ltx N} \trace \left ( \EE \left [ \Xm_c^\herm (f) \Xm_c(f) \right ] \right )
%& = &  \frac{1}{\Ltx N} \trace \left ( \EE \left [  \Vm_c (f)  \Um_c^\herm (f) \Um_c(f) \Vm_c^\herm(f) \right ] \right ) \nonumber \\
%& = &  \frac{\Ltx N }{S \Ltx N} \trace \left ( \EE \left [ \Vm_c^\herm(f)  \Vm_c (f)  \right ] \right ) \nonumber \\
%& = &  \frac{SN}{S}  = N,
%\end{eqnarray}
$\frac{1}{\Ltx N} \trace \left ( \EE \left [ \Xm_c^\herm (f) \Xm_c(f) \right ] \right ) = N$, as desired.

Recalling the definition of $\Dc(f)$, the received signal for user $k$ at location $x + c : x \in \Xc$ is given by
\begin{equation} \label{eq:received-signal}
\yv_{k,c}(f;x) = \sum_{c' \in \Dc(f)} \Um_{c'}(f) \Vm^\herm_{c'}(f) \underline{\hv}_{k,c,c'}(f;x) + \zv_{k,c}(f;x)
\end{equation}
where $\zv_{k,c}(f;x) \sim \Cc\Nc\left(\zerov, \frac{1}{F} \Id_{\Ltx N}\right)$.
Notice that a scheme using frequency reuse $F > 1$ transmits with total cluster power $N$ over a fraction
$1/F$ of the whole system bandwidth. This is taken into account by letting the noise variance
per component be equal  to $1/F$, in the signal model (\ref{eq:received-signal}).

By construction, the encoded data symbols  for user $k$ at location $x + c : x \in \Xc$, are the entries of
the $k$-th column  of $\Um_{c}(f)$. The columns $k' \neq k$ of  $\Um_{c}(f)$ form the intra-cluster (multiuser)
interference for user $k$.  All other signals $\Um_{c'}(f)$, with $c' \in \Dc(f), c' \neq c$, form the Inter-Cluster Interference (ICI).
As seen in Section \ref{sec:precoding}, intra-cluster interference and ICI are handled by a combination
of beamforming and frequency reuse.

%%%%%%%%%%%%%%%%%%%%%%%%%%%%%%%%%%%%%%%%%%%%%%%%%%%%%%%
\section{Uplink training and channel estimation} \label{sec:training}

The CSIT is obtained on a per-slot basis,
by letting all the scheduled (i.e., active) users in the slot sent pilot signals over
the $\Ltr N$ dimensions dedicated to uplink training.\footnote{As done in \cite{Marzetta-TWC10}, also
our analysis is slightly optimistic since it only accounts for the overhead and degradation
due to uplink noisy channel estimation, while it assumes genie-aided overhead-free
``dedicated training'' to support coherent detection during data-transmission.
As shown in \cite{Caire-Jindal-Kobayashi-Ravindran-TIT10}, the effect of noisy dedicated training is minor
relatively to the CSIT estimation error.}
We fix $\{\Xc,\Cc\}$ and focus on the $SN$ active users in the groups $\Xc + c : c \in \Dc(f)$.
These users must send  $S N$ orthogonal pilot signals to allow channel estimation at their corresponding serving
clusters $\Cc + c : c \in \Dc(f)$.

%%%%%%%%%%%%%%%%%%%%%%%%%%%%%%%%%%%%%%%%%%%%%%%%%%
\subsection{Pilot reuse scheme} \label{subsec:pilot}

Let $\Ltr = Q S$, where $Q \geq 1$ is an integer {\em pilot reuse factor} that can be optimized for each $\{\Xc,\Cc\}$.
Let $\Phim \in \CC^{QSN \times QSN}$ be a scaled unitary matrix, such that $\Phim^\herm \Phim = \alpha_{\rm ul} QSN \Id_{QSN}$,
where $\alpha_{\rm ul}$ denotes the uplink  transmit power per user during the training phase.
The columns of $\Phim$ are partitioned into $Q$
disjoint blocks of size $SN$ columns each, denoted by $\Phim_{0},\ldots,\Phim_{Q-1}$ and referred to as
{\em training codebooks}. These are assigned to the groups in a periodic fashion,
such that the same training codebook $\Phim_q$ is reused every $Q$-th groups $\Xc + c : c \in \Dc(f)$.
For later use, we let $q(c) \in \{0,\ldots, Q-1\}$ denote the index of the training
codebook allocated to group  $\Xc+c$, and define $\Pc(q, f) = \{ c \in \Dc(f) : q(c) = q\}$ as the set of clusters active on subband $f$ and using
training codebook $q$. Pilot reuse is akin frequency reuse, but in general $Q$ and $F$ may be different in order to
allow for additional flexibility in the system optimization.

\begin{figure}
  \centering
  \includegraphics[width=12cm]{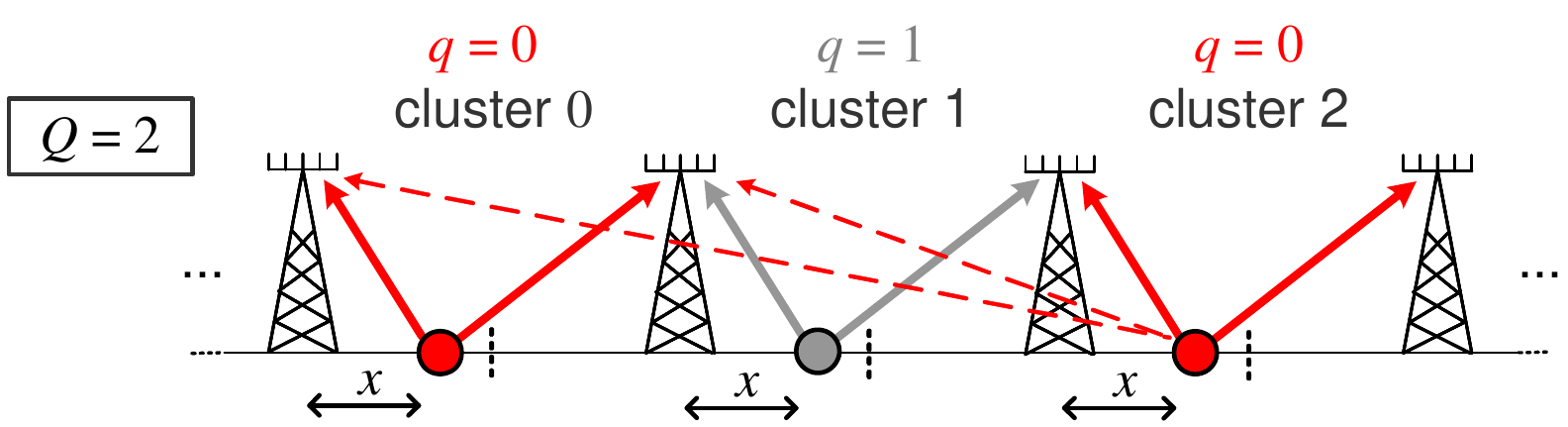}
  \caption{Pilot reuse and contamination for $C=2$, $F=1$, and $Q=2$. The dashed lines show the contamination from
  a user sharing the same pilot signal, in another cluster.}
  \label{fig:pilot-conta}
\end{figure}

\begin{example} \label{pilot-reuse-example}
In the 1-dimensional layout with $C = 2$, $\Cc = \{0,1\}$ and $\Xc = \{x, 1 - x \}$ we may have
$F = 1$ (i.e., each cluster is active on the whole system bandwidth) and $Q = 2$ (i.e., two mutually orthogonal training codebooks
are used alternately, such that the same set of uplink pilot signals is reused in every other cluster,
as shown in Fig.~\ref{fig:pilot-conta}).
\hfill $\lozenge$
\end{example}

%%%%%%%%%%%%%%%%%%%%%%%%%%%%%%%%%%%%%%%%%%
\subsection{MMSE channel estimation and pilot contamination} \label{subsec:mmse}

The uplink signal received by the $CMN$ antennas of cluster
$\Cc + c : c \in \Dc(f)$, during the training phase, is given by
\begin{equation} \label{eq:received-signal-training-TDD}
\Ym_c(f) = \sum_{c'\in \Dc(f)} \Phim_{q(c')} \underline{\Hm}_{c',c}^\herm (f; \Xc) + \Zm_c(f).
\end{equation}
Because of TDD reciprocity, the {\em uplink} channel matrix $\underline{\Hm}_{c',c}(f; \Xc) \in \CC^{CMN \times SN}$ contains
the {\em downlink} channels $\underline{\hv}_{k,c',c}(f;x)$ arranged by columns, for all active users $k = 1, \ldots, SN/m$
at all  locations $x + c' :  x \in \Xc$. In (\ref{eq:received-signal-training-TDD}), $\Zm_c(f) \in \CC^{\Ltr \times CMN}$ denotes the uplink AWGN with components $\sim \Cc\Nc(0,1)$.
The goal of the uplink training phase is to provide to each cluster $\Cc+c$ an estimate of the channel vectors $\underline{\hv}_{k,c,c}(f;x)$ for all
the active users in the corresponding served group $\Xc + c$.

By projecting $\Ym_c(f)$  onto the column of $\Phim_{q(c)}$ associated to user $k$ at location
$x+c : x \in \Xc$ and dividing by $\alpha_{\rm ul} QSN$, the relevant observation for estimating the
$\underline{\hv}_{k,c,c}(f;x)$ is given by
\begin{equation} \label{eq:received-signal-ch-est-TDD}
\underline{\rv}_{k,c}(f; x) =  \sum_{c' \in \Pc(q(c),f)} \underline{\hv}_{k,c',c}(f; x) +  \underline{\nv}_{k,c}(f)
\end{equation}
where $\underline{\nv}_{k,c}(f) \sim \Cc\Nc(\zerov, (\alpha^{\rm ul} Q S N)^{-1} \Id_{CMN})$.
For any $c' \in \Pc(q(c),f)$, the MMSE estimate of $\underline{\hv}_{k,c',c}(f;x)$ from $\underline{\rv}_{k,c}(f; x)$ is obtained as
\begin{equation} \label{mmse-estimator}
\widehat{\underline{\hv}}_{k,c',c}(f;x) = \Gm_{c',c}(x) \left [ (\alpha^{\rm ul} Q S)^{-1} \Id_{CMN} +
\sum_{c'' \in \Pc(q(c),f)} \Gm_{c'',c}(x) \right ]^{-1} \underline{\rv}_{k,c}(f; x)
\end{equation}
Invoking the well-known MMSE decomposition, we can write
\begin{equation} \label{eq:MMSE-decomp}
\underline{\hv}_{k,c',c}(f;x) = \widehat{\underline{\hv}}_{k,c',c}(f;x) + \underline{\ev}_{k,c',c}(f;x),
\end{equation}
where the channel estimate $\widehat{\underline{\hv}}_{k,c',c}(f;x)$ and the
error vector $\underline{\ev}_{k,c',c}(f;x)$ are zero-mean uncorrelated jointly complex circularly symmetric
Gaussian vectors (and therefore statistically independent due to joint Gaussianity).
After some straightforward algebra (omitted for brevity), we obtain the covariance matrices
$\EE[ \widehat{\underline{\hv}}_{k,c',c}(f;x) \widehat{\underline{\hv}}^\herm_{k,c',c}(f;x) ] = \frac{1}{N} \Xim_{c',c}(x)$ and
$\EE[\underline{\ev}_{k,c',c}(f;x) \underline{\ev}^\herm_{k,c',c}(f;x)] = \frac{1}{N} \Sigmam_{c',c}(x)$, where
$\Xim_{c',c}(x) = \diag\left ( \xi_{c',c,b}(f;x) \Id_{MN} : b \in \Cc \right )$
and $\Sigmam_{c',c}(x) = \diag\left ( \sigma_{c',c,b}(f;x) \Id_{MN} : b \in \Cc \right )$,
and where we define
\begin{eqnarray}
\sigma_{c',c,b}(f;x) & = & \frac{g(x+c',b+c)}{1+\gamma_{c',c,b}(f;x)} \label{eq:estimation-mmse} \\
\xi_{c',c,b}(f;x) & = & g(x+c',b+c) - \sigma_{c',c,b}(f;x)
  = \frac{g(x+c',b+c)}{1+\gamma_{c',c,b}(f;x)^{-1}} \label{xixi} \label{eq:eff-pathloss}
\end{eqnarray}
with
\begin{equation} \label{gamma}
\gamma_{c',c,b}(f;x) = \frac{g(x+c',b+c)}{(\alpha_{\rm ul} QS)^{-1}  + \sum_{c'' \in \Pc(q(c),f) \backslash c'} g(x+c'',b+c)}
\end{equation}
The desired channel estimate at cluster $\Cc+c$ is given by $\widehat{\underline{\hv}}_{k,c,c}(f;x)$,
obtained by letting  $c' = c$ in (\ref{mmse-estimator}) -- (\ref{gamma}).
Notice that the training phase observation $\underline{\rv}_{k,c}(f;x)$ in (\ref{eq:received-signal-ch-est-TDD}) contains
the superposition of all the channel vectors $\underline{\hv}_{k,c',c}(f;x)$ of the users $k$ at location $x+c' : x \in \Xc$,
for all $c' \in \Pc(q(c),f)$, i.e., sharing the same pilot signal.
This is the so-called {\em pilot contamination effect}, which is a major
limiting factor in the performance of TDD systems \cite{Jose-Ashikhmin-Marzetta-Vishwanath-TWC11,Marzetta-TWC10}.
Because of pilot contamination, the MMSE estimate  $\widehat{\underline{\hv}}_{k,c,c}(f;x)$ is
{\em correlated} with the channels $\underline{\hv}_{k,c',c}(f;x)$, for all $c' \in \Pc(q(c),f)$.

Next, we express the channel vector $\underline{\hv}_{k,c',c}(f;x)$ for  $c' \in \Pc(q(c),f)$ in terms of the estimate
$\widehat{\underline{\hv}}_{k,c,c}(f;x)$ and a component independent of $\widehat{\underline{\hv}}_{k,c,c}(f;x)$.
This decomposition is useful to proof the main results of Theorems 1, 2 and 3 in Section \ref{sec:efficiency}
%
%will be used in the proofs of the main analysis results
%in Appendices \ref{appen:bin-rate-mrt} and \ref{appen:bin-rate-zfbf}
and it is the key to understand qualitatively the pilot contamination effect.
From (\ref{mmse-estimator}), and since $\Sigmam_{c,c}(x)$ is invertible, we have
\begin{eqnarray} \label{nonsomica}
\widehat{\underline{\hv}}_{k,c',c}(f;x) & = & \EE\left [\underline{\hv}_{k,c',c}(f;x) | \underline{\rv}_{k,c}(f;x) \right ] \nonumber \\
%& = & \EE \left [\underline{\hv}_{k,c',c}(f;x) | \widehat{\underline{\hv}}_{k,c,c}(f;x) \right ] \nonumber \\
& = & \Gm_{c',c}(x) \Gm_{c,c}^{-1} (x)   \widehat{\underline{\hv}}_{k,c,c}(f;x).
\end{eqnarray}
Using (\ref{eq:MMSE-decomp}) into (\ref{nonsomica}), the channel vector $\underline{\hv}_{k,c',c}(f;x)$ from the antennas of cluster $\Cc + c$ to the
{\em unintended} user $k$ at location $x + c' : x \in \Xc$ can be written as
\begin{equation} \label{eq:MMSE-decomp1}
\underline{\hv}_{k,c',c}(f;x) = \Gm_{c',c}(x) \Gm_{c,c}^{-1} (x)   \widehat{\underline{\hv}}_{k,c,c}(f;x) + \underline{\ev}_{k,c',c}(f;x).
\end{equation}
Joint Gaussianity,  the mutual orthogonality of $\widehat{\underline{\hv}}_{k,c',c}(f;x)$ and $\underline{\ev}_{k,c',c}(f;x)$ and the fact that
all covariance matrices are diagonal imply
% the MMSE estimator (\ref{mmse-estimator}) decouples into componentwise individual scalar estimators,
that $\widehat{\underline{\hv}}_{k,c,c}(f;x)$ and $\underline{\ev}_{k,c',c} (f;x)$ are
mutually independent.

%%%%%%%%%%%%%%%%%%%%%%%%%%%%%%%%%%%%%%%%%
%%%%%%%%%%%%%%%%%%%%%%%%%%%%%%%%%%%%%%%%% PILOT CONTAMINATION
As anticipated before, (\ref{eq:MMSE-decomp1}) reveals qualitatively the pilot contamination effect.
With LSUBF, as in \cite{Marzetta-TWC10}, cluster $\Cc + c$ serves user $k$ at location $x + c$ with beamforming vector
$\widehat{\underline{\hv}}_{k,c,c}(f;x)/\|\widehat{\underline{\hv}}_{k,c,c}(f;x)\|$, which is strongly correlated with the channel vector
$\underline{\hv}_{k,c',c}(f;x)$ towards the unintended user $k$ at location $x + c'$, sharing the same pilot signal.
It follows that some constant amount of interfering power, that does not vanish with $N \rightarrow \infty$, is sent in the ``spatial direction''
of this user, leading to an interference limited system, as exactly quantified by Theorem 1 in Section \ref{sec:efficiency}.
For the family of LZFBF schemes considered in this work, the pilot contamination effect is less intuitive, and it is precisely
quantified by Theorems 2 and 3 in Section \ref{sec:efficiency}.

%in the uplink. This requires that at least $S$ dimensions out of the $L$ dimensions per coherence block are used by the pilot signals.
%Eventually, the same pilot signals are reused in other cells, and the CSIT is ``contaminated'' by the interference from the pilot signals reused
%in other cells. Hence,  the contaminated beamforming vectors focus some transmit power in the direction  of the unintended users
%in the other cells sharing the same pilot signals. This effect is particularly evident for users at the cell edge.

%%%%%%%%%%%%%%%%%%%%%%%%%%%%%%%%%%%%%%%%%%%%%%%%%%%%%%%
\section{MU-MIMO Precoders and Achievable Rates} \label{sec:precoding}

In the family of network-MIMO schemes considered in this work, the beamforming matrix $\Vm_c(f)$ is calculated as a function of the estimated
channel matrix $\widehat{\underline{\Hm}}_{c,c}(f; \Xc)$. The schemes differ by the type of beamforming employed.
In particular, we consider LZFBF where any active user $k$ at location $x + c: x \in \Xc$, imposes ZF constraints on $J \geq 0$ clusters.
A ZF constraint consists of the set of linear equations
\begin{equation} \label{ZFconstr}
\vv_{j,c'}^\herm (f;x')  \widehat{\underline{\hv}}_{k,c,c'}(f;x) =  0, \;\;\; \forall \;\; (j,x',c') \neq (k,x,c)
\end{equation}
where $\vv_{j,c'} (f;x')$ denotes the column of $\Vm_{c'}(f)$ corresponding to user $j$ at location $x' + c' : x' \in \Xc$.

%%%%%%%%%%%%%%%%%%%%%%%%%%%%%%%%%%%%%%%%%%%%%%%%%%%%%%%
\subsection{Beamforming} \label{precoding-types}

Next we provide expressions for the cluster precoders for different choice of the parameter $J$.

{\bf Case $J = 0$:} In this case no ZF constraints are imposed. Hence, we have
\begin{equation} \label{lsubf-precoding}
\Vm_{c}(f) = {\rm UNorm}\left \{  \widehat{\Hm}_{c,c}(f; \Xc) \right\}
\end{equation}
where the operation UNorm$\{\cdot\}$ indicates a scaling of the columns of the matrix argument such that they have unit norm.
It is immediate to see that (\ref{lsubf-precoding}) coincides with the Linear Single-User Beamforming (LSUBF) considered in \cite{Marzetta-TWC10}.
%, where
%the beamforming vector $\vv_{k,c}(f;x)$ is a scaled version of the corresponding estimated channel vector $\widehat{\underline{\hv}}_{k,c,c}(f;x)$.

{\bf Case $J = 1$:} In this case any active user imposes ZF constraints
on its own serving cluster. This yields the classical single-cluster LZFBF, for which
\begin{equation}  \label{zfclassic}
\Vm_{c}(f) = {\rm UNorm}\left \{  \widehat{\Hm}_{c,c}^+(f; \Xc) \right\},
\end{equation}
where
\begin{equation} \label{eq:def-pinv}
\Mm^+ = \Mm \left[ \Mm^\herm \Mm \right]^{-1}
\end{equation}
denotes the Moore-Penrose pseudo-inverse of the full column-rank matrix $\Mm$.
It follows that $\vv_{k,c}(f;x)$ is orthogonal to the estimated channels $\widehat{\underline{\hv}}_{j,c,c}(f;x')$
for all other active users $(j,x') \neq (k,x)$ in the same cluster $\Cc + c$, i.e., ZF is used to tackle intra-cluster
interference,  but nothing is done with respect to ICI.
%The large-system asymptotic analysis of LZFBF with joint transmission
%from clusters of BSs, distance dependent pathloss and channel estimation errors has been recently provided in
%\cite{Huh-Tulino-Caire-TITsubmit}.

{\bf Case $J > 1$:} In this case, beyond the ZF constraints imposed to the serving cluster, each user imposes additional ZF constraints to $J - 1$ neighboring clusters
in order to mitigate the ICI. Mitigating ICI through the beamforming design provides an alternative approach to frequency reuse and,
in general,  might be used jointly with frequency reuse.  Let's focus on cluster $\Cc+c$.
This is subject to ZF constraints imposed by its own users (i.e., users in group $\Xc + c$),
as well as by some users at some locations $x' + c' : x' \in \Xc$ for $J-1$ neighboring clusters $c' \neq c$.
In order to enable such constraints,  the $c$-th cluster controller must be able to estimate
the channels of these out-of-cluster users.  This can be done if these users employ training codebooks with
indices $q \neq q(c)$. In particular, $J > 1$ can be used only if the pilot reuse factor $Q$ is larger than 1.
In some cases, only the channel subvectors to the nearest BS in the cluster can be effectively estimated, since
there are other users sharing the same pilot signal that are received with a stronger path coefficient.
Then, the channel subvectors  that  cannot be estimated are treated as zero.
Since these schemes are complicated to explain in full generality, we shall illustrate
two specific examples, the generalization of which is cumbersome but conceptually straightforward,
and can be worked out by the reader if interested in other specific cases.

\begin{figure}
  \centering
  \subfigure[$J=Q$]{
  \includegraphics[width=10cm]{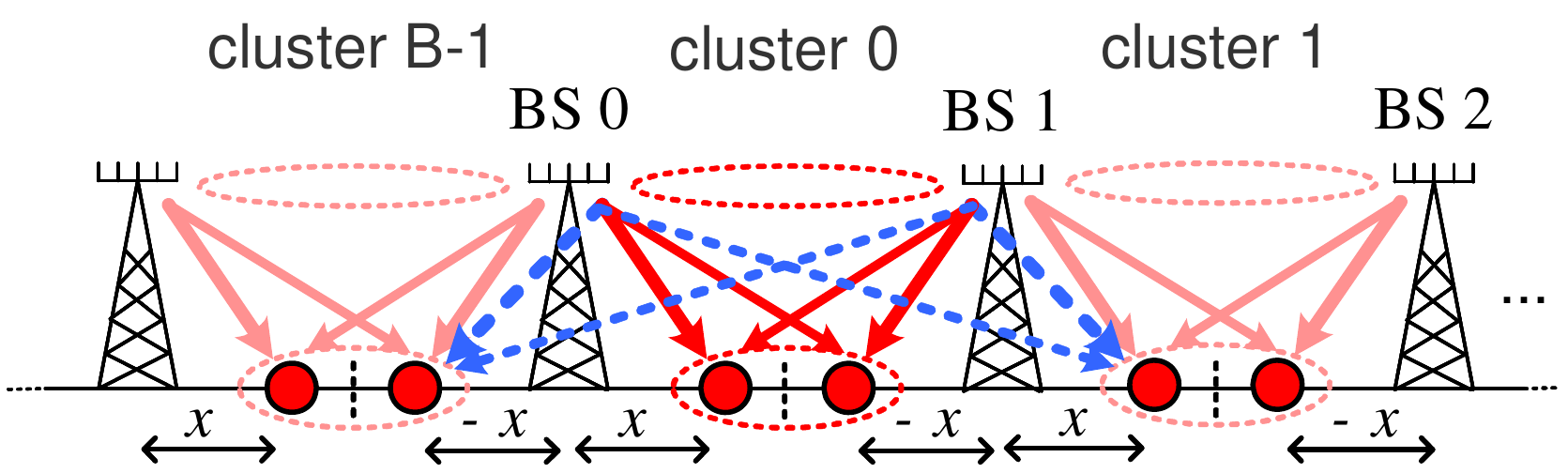}
  \label{subfig:zfcluster-caseb}}
  \subfigure[$J=C(Q-1)+1$]{
  \includegraphics[width=10cm]{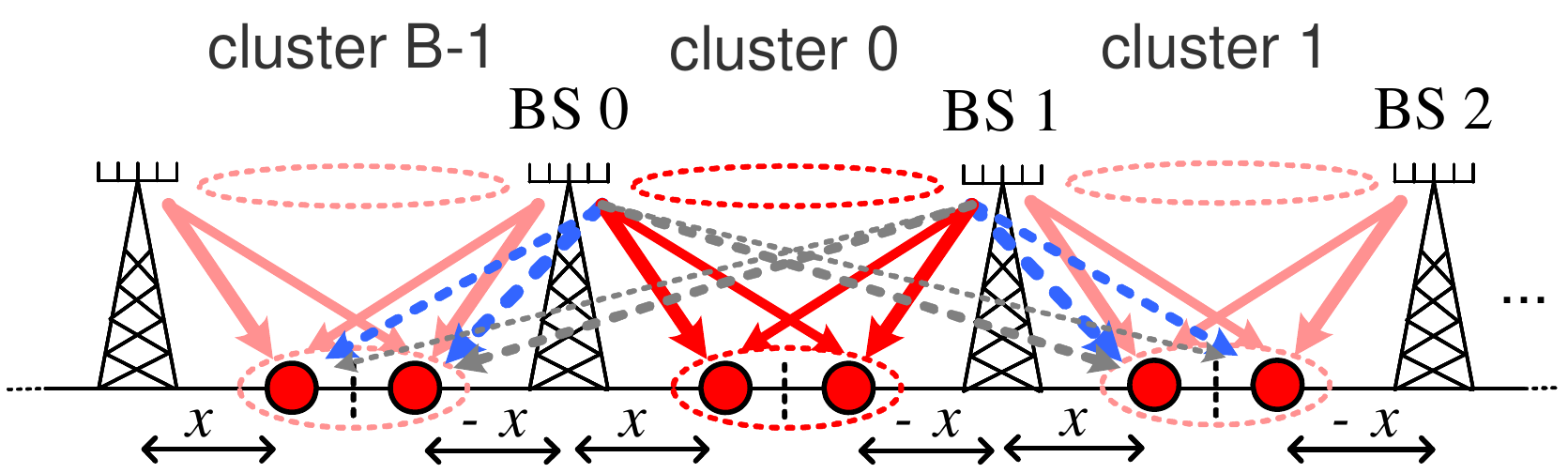}
  \label{subfig:zfcluster-casec}}
  \caption{Two cases of a precoding scheme for $C=2$, $F=1$, and $Q=2$, with $J = Q$ (a) and $J = C(Q-1)+1$ (b). The dashed lines indicate
  the channel vectors to out-of-cluster users for which a ZF constraint is imposed. In Figure (b), the light-shaded dashed lines indicate the channel vectors assumed
  zero in the beamforming calculation.
  }
  \label{fig:zfclster-casebc}
\end{figure}

\begin{example} \label{JequalQ}
Consider Fig.~\ref{subfig:zfcluster-caseb}, illustrating a 1-dimensional system
with $\Cc = \{0,1\}$, $\Xc = \{x, 1 - x\}$, $F = 1$ and $Q = 2$.
The beamforming matrix of each cluster $c$ satisfies ZF constraints for its own served users
and for the users in the $m = 2$ locations  at minimum distance in the nearest neighbor clusters, $c-1$ and $c+1$.
These locations collectively use distinct columns of  the training codebooks $q \neq q(c)$. In the specific example, the reference cluster $c = 0$ uses
training codebook $\Phim_0$, and the nearest locations on the left and on the right of cluster 0 use the first $SN/2$ columns and the second $SN/2$ columns
of the other training codebook $\Phim_1$, respectively. Hence, cluster 0 controller can estimate all the channels of its own active users,
at locations $x, 1 - x$, and of the users in adjacent locations $-x$ and $1+x$, as shown in the figure.
The beamforming matrix in the case of Fig.~\ref{subfig:zfcluster-caseb} is obtained as follows. Define
\begin{equation} \label{ziofa-caseb}
\Mm_c(f;\Xc) =
\Big [
\underbrace{\widehat{\underline{\Hm}}_{c,c}(f;\Xc)}_{2MN\times SN} \Big |
\underbrace{\widehat{\underline{\Hm}}_{c-1,c}(f;\{1-x\})}_{2MN\times SN/2}  \Big |
\underbrace{\widehat{\underline{\Hm}}_{c+1,c}(f;\{x\})}_{2MN\times SN/2} \Big ]
\end{equation}
be the matrix of dimension $2MN \times 2SN$ of all estimated channels at cluster controller $c$, where the first block
corresponds to the desired active users and the remaining blocks
correspond to users in the adjacent clusters for which a ZF constraint is imposed.  Then,
\begin{equation}  \label{zf-selection}
\Vm_{c}(f) = \left [ {\rm UNorm}\left \{  \Mm_{c}^+(f;\Xc) \right\} \right ]_{k = 1}^{SN}
\end{equation}
where $[\cdot]_n^m$ extracts the columns from $n$ to $m$ of the matrix argument.  This scheme can be generalized to $J = Q$, where each
cluster $c$ satisfies ZF constraints for the desired $SN$ active users in its own cluster and
for a total of $(J - 1)SN$ additional users in the nearest location of
neighboring clusters. \hfill $\lozenge$
\end{example}

\begin{example} \label{JequalCQ-1+1}
Consider Fig.~\ref{subfig:zfcluster-casec}, illustrating the same 1-dimensional system as in Example \ref{JequalQ} with a different
beamforming design. In this case,
the beamforming matrix of each cluster $c$ satisfies ZF constraints for its own served users
and {\em all} the users in the nearest neighbor clusters.
However,  some of these users share the same columns of the training codebooks $q \neq q(c)$.
In the specific example, the reference cluster $c = 0$ uses training codebook $\Phim_0$, and the clusters to the left
and to the right the other training codebook $\Phim_1$.
Users at location $-1 + x$ use the same pilot signals of users at location $1 + x$,
and users  at location $-x$ use the same pilot signals of users at location $2-x$.
Then the 0-th cluster controller assumes that the channel coefficients for  BSs at larger distance are equal to zero.
In the example,  for locations $-1+x$ and $-x$, only the subvector of dimension $MN$
corresponding to the antennas of  BS 0 is estimated, while the remaining subvector to BS 1 is treated as zero.
Similarly,  for locations $1 + x$ and $2 - x$ only the subvector
of dimension $MN$ corresponding to the antennas of BS 1 is estimated, while the remaining subvector to BS 0
is treated as zero.

The beamforming matrix corresponding to the scheme of Fig.~\ref{subfig:zfcluster-casec} is obtained as follows. Define
\begin{equation} \label{ziofa-casec}
\Mm_c(f;\Xc) =
\Big [ \underbrace{\widehat{\underline{\Hm}}_{c,c}(f;\Xc)}_{2MN\times SN} \Big |
\underbrace{\widehat{\underline{\Hm}}_{c-1,c}(f;\Xc)}_{2MN\times SN} \odot \Lm \Big |
\underbrace{\widehat{\underline{\Hm}}_{c+1,c}(f;\Xc)}_{2MN\times SN/2} \odot \Rm \Big ]
\end{equation}
be the matrix of dimension $2MN \times 3SN$ of all estimated channels at cluster controller $c$,
where $\odot$ indicates elementwise product and where
\[
\Lm = \left [ \begin{array}{c} \onev_{MN \times SN} \\ \zerov_{MN\times SN} \end{array} \right ] , \;\;\;\;
\Rm = \left [ \begin{array}{c} \zerov_{MN \times SN} \\ \onev_{MN\times SN} \end{array} \right ]  \]
are masking matrices that null out the subvectors of the channels that are treated as zero in the beamforming design.
Then, $\Vm_{c}(f)$ is again given by (\ref{zf-selection}) although in this case $\Mm_c(f;\Xc)$ is given by (\ref{ziofa-casec}) instead of
(\ref{ziofa-caseb}).
This scheme can be generalized to $J = C(Q-1)+1$, where each cluster $c$ satisfies ZF constraints for the desired
$SN$ active users in its own cluster and  for a total of $(J - 1)SN$ additional users in the
neighboring clusters, with some channel sub-vectors set to zero. \hfill $\lozenge$
\end{example}

%%%%%%%%%%%%%%%%%%%%%%%%%%%%%%%%%%%%%%%%%%
\subsection{Achievable group spectral efficiency}  \label{sec:efficiency}

Letting $R_{k,c}^{(N)}(f; x)$ denote the spectral efficiency (in bit/s/Hz) of user $k$ at location $x+c: x\in \Xc$, served by cluster $c$ according to
a scheme as defined above,  we define the group spectral efficiency of bin $v(\Xc)$ as
\begin{equation} \label{norm-bin-speff}
R_{\Xc, \Cc}(F, C, J) = \frac{1}{FBN} \sum_{f = 0}^{F-1} \sum_{c \in \Lambda_{\rm bs} \cap \Vc} \sum_{x\in \Xc} \sum_{k=1}^{SN/m} R^{(N)}_{k,c}(f;x)
\end{equation}
In Appendices \ref{appen:bin-rate-mrt} and \ref{appen:bin-rate-zfbf}, we prove the following results.

%%%%%%%%%%%%%%%%%%%%%%%%%%%%%%%%%%%%%%%%%%%%%%%%%%%%%%%%%%%%%%%
%%%%%%%%%%%%%%%%%%%%%%%%%%%%%%%%%%%%%%%%%%%%%%%%%%%%%%%%%%%%%%%

\begin{theorem} \label{J0thm}
For given sets $\Xc, \Cc$,
and system parameters $F, S$ and $Q$,
%frequency reuse factor $F$, downlink loading factor $S$ and pilot reuse factor $Q$,
in the limit of $N \rightarrow \infty$, the following group spectral efficiency of bin $v(\Xc)$ is  achievable with LSUBF precoding ($J = 0$):
\begin{eqnarray} \label{eq:bin-rate-lsubf}
R_{\Xc, \Cc}(F,C,J=0)
%& = & \lim_{N \rightarrow \infty} \frac{1}{NF} \sum_{x \in \Xc} \sum_{k=1}^{SN/m} R^{(N)}_{k,0}(x) \nonumber \\
& = & \frac{S}{mF} \sum_{x\in \Xc} \log \left ( 1 + \frac{\frac{CM}{S} \underline{\xi}_{0,0}(x)}{\frac{1}{F}
+ \eta(x)  +  \frac{CM}{S}  \zeta(x) } \right ),
\end{eqnarray}
where\footnote{In (\ref{etax}) and (\ref{zetax}) it is assumed, without loss of generality,  that cluster $c = 0$ uses subband $f = 0$ and training codebook $q = 0$.}
\begin{equation} \label{etax}
\eta(x) = \frac{1}{mC}   \sum_{x' \in \Xc}  \sum_{b \in \Cc}  \sum_{c \in \Dc(0)}   \frac{\xi_{c,c,b}(x') g(x,c+b)}{\underline{\xi}_{c,c}(x')}
\end{equation}
and
\begin{equation} \label{zetax}
\zeta(x) = \sum_{c \in \Pc(0,0) \backslash 0}  \frac{1}{\underline{\xi}_{c,c}(x)} \left ( \frac{1}{C} \sum_{b \in \Cc} \frac{g(x,c+b)}{g(x,b)} \xi_{c,c,b}(x) \right )^2,
\end{equation}
with
\begin{equation} \label{xibar}
\underline{\xi}_{c,c}(x) = \frac{1}{C} \sum_{b\in \Cc} \xi_{c,c,b}(x),
\end{equation}
are coefficients that depend uniquely on the system geometry, frequency and pilot reuse, but are independent of the loading factor
$S$ and of the BS antenna factor $M$.
%Notice that in order to obtain (\ref{eq:bin-rate-lsubf}) we have used the fact that, by construction, the group spectral efficinecy
%is symmetric with respect to the cluster index $c$ and the frequency subband $f$.
\hfill \IEEEQED
\end{theorem}

As a corollary of Theorem \ref{J0thm}, we can recover the result of \cite{Marzetta-TWC10}. It is sufficient to let $M \rightarrow \infty$ in
(\ref{eq:bin-rate-lsubf}) and obtain the regime of infinite number of BS antennas per active user. Particularizing this for
fixed $S$, $C = 1$, and $Q=1$ as in \cite{Marzetta-TWC10}, the group spectral efficiency becomes
\begin{eqnarray} \label{binrate-lsubf-Marzetta}
\lim_{M \rightarrow \infty} R_{\Xc, \{0\}}(F,1,0)
  = \frac{S}{mF} \sum_{x \in \Xc} \log \left ( 1 + \frac{g(x,0)^2}
  {\sum_{c \in \Pc(0,0) \backslash 0} g(x,c)^2} \right )
\end{eqnarray}
As observed in \cite{Marzetta-TWC10},  in this regime the system spectral efficiency is uniquely limited by
the ICI due to pilot contamination.

The next result yields the achievable group spectral efficiency of LZFBF in the case of single-cell processing (i.e., for $C = 1$).
We define $\Ec(x)$ as the set of  $J-1$  clusters $c \neq 0$ with centers closest  to $x \in \Xc$
(if $J = 1$ then  $\Ec(x) = \emptyset$). Then, we have:

\begin{theorem} \label{Jgeq1C1thm}
For given set $\Xc$, $C = 1$ (i.e., $\Cc = \{0\}$),
and system parameters $F, S$ and $Q$,
%
%frequency reuse factor $F$, downlink loading factor $S$ and pilot reuse factor $Q$,
in the limit of $N \rightarrow \infty$, the following group spectral efficiency of bin $v(\Xc)$ is  achievable with LZFBF precoding  ($J \geq 1$):
\begin{align} \label{eq:bin-rate-zfbf}
R_{\Xc, \{0\}}(F,1,J\geq 1)
%\;& = \; \lim_{N \rightarrow \infty} \frac{1}{NF}
%  \sum_{x \in \Xc} \sum_{k=1}^{SN/m} R^{(N)}_{k,0}(x) \nonumber \\
& =\; \frac{S}{mF} \sum_{x \in \Xc} \log \left( 1 + \frac{\frac{M-JS}{S} \xi_{0,0,0}(x)}{\frac{1}{F} + \alpha(x) + \frac{M-JS}{S} \beta(x) } \right)
\end{align}
where
\begin{equation} \label{alphax}
\alpha(x) =  \sum_{c \in \Pc(0,0) \cup \Ec(x)} \sigma_{0,c,0}(x) + \sum_{c \in \Dc(0) - \Pc(0,0) - \Ec(x)} g(x,c)
\end{equation}
and
\begin{equation} \label{betax}
\beta(x) = \sum_{c \in \Pc(0,0) \backslash 0}
 \left (  \frac{g(x,c)}{g(x,0)} \right )^2 \xi_{c,c,0}(x)
\end{equation}
are coefficients that depend uniquely on the system geometry, frequency and pilot reuse, but are independent of the loading factor $S$ and of
the BS antenna factor $M$.
%Notice that in order to obtain (\ref{eq:bin-rate-lsubf}) we have used the fact that, by construction, the group spectral efficinecy
%is symmetric with respect to the cluster index $c$ and the frequency subband $f$.
\hfill \IEEEQED
\end{theorem}

In passing, we notice that the limit of (\ref{eq:bin-rate-zfbf})
for $M \rightarrow \infty$, coincides with (\ref{binrate-lsubf-Marzetta}).
Therefore, as observed in \cite{Marzetta-TWC10}, in the ``Massive MIMO'' regime LZFBF yields no advantage over
the simpler LSUBF.

The case of LZFBF with multicell processing ($C > 1$) needs some more notation.
First, as illustrated in Examples \ref{JequalQ} and \ref{JequalCQ-1+1}, we consider the cases
$J = 1$, $J = Q$ and $J = C(Q-1)+1$, referred to as cases (a), (b) and (c), respectively, for the sake of brevity.
In case (c) it is useful to define $b(x,c) = \arg\min\{ d_{\Lambda}(x,c+b) : b \in \Cc\}$, i.e.,
the closest BS to location $x \in \Xc$  in cluster $c \in \Ec(x)$.
For $C > 1$, an exact asymptotic ICI power expression cannot be found due to the complicated statistical dependence
of beamforming vectors and channel vectors due to pilot contamination.
However, the following result yields an achievable rate based on an upper bound on the ICI power  (see details in Appendix \ref{appen:bin-rate-zfbf}):

\begin{theorem} \label{Jgeq1Cgeq1thm}
For given sets $\Xc, \Cc$ with $C > 1$,  and system parameters $F, S$ and $Q$,
%For given $\Xc$ and $\Cc$ with $C > 1$, frequency reuse factor $F$, downlink loading factor $S$ and pilot reuse factor $Q$,
in the limit of $N \rightarrow \infty$, the following group spectral efficiency of bin $v(\Xc)$ is  achievable with LZFBF precoding  ($J \geq 1$):
\begin{align} \label{eq:bin-rate-zfbf-lb}
R_{\Xc, \Cc}(F,C,J\geq 1)
%\;& = \; \lim_{N \rightarrow \infty} \frac{1}{NF}
%  \sum_{x \in \Xc} \sum_{k=1}^{SN/m} R^{(N)}_{k,0}(x) \nonumber \\
& =\; \frac{S}{mF} \sum_{x\in \Xc} \log \left( 1 + \frac{\frac{CM-JS}{S} \underline{\xi}_{0,0}(x)}{\frac{1}{F} + \underline{\alpha}(x) + \frac{CM}{S} \underline{\beta}(x) } \right)
\end{align}
where
\begin{equation} \label{underalphax}
\underline{\alpha}(x) =
\left\{ \begin{array}{l}
\displaystyle{ \sum_{c \in \Ec(x) \cup \{0\}} \underline{\sigma}_{0,c}(x) + \sum_{c \in \Dc(0)\backslash 0  - \Ec(x)} \underline{g}_{0,c}(x)}, \qquad \mbox{in cases (a) and (b)}, \\
  \\
\displaystyle{ \underline{\sigma}_{0,0}(x) + \sum_{c \in \Dc(0)\backslash 0  - \Ec(x)} \underline{g}_{0,c}(x)}  +  \\
  + \displaystyle{  \frac{1}{C} \sum_{c \in \Ec(x)} \left( \sigma_{0,c,b(x,c)}(x)     +                      \sum_{b\in \Cc\backslash b(x,c)} g(x,c+b)   \right)}
  \qquad \mbox{in case (c),}  \end{array} \right.
\end{equation}
and
\begin{equation} \label{underbetax}
\underline{\beta}(x) = \sum_{c \in \Pc(0,0) \backslash 0} \frac{1}{C} \sum_{b\in \Cc}
  \left ( \frac{g(x,c+b)}{g(x,b)}\right)^2 \xi_{c,c,b}(x),
\end{equation}
with
\begin{equation} \label{gg}
\underline{g}_{0,c}(x)  = \frac{1}{C} \sum_{b\in \Cc} g(x,c + b), \;\;\;\; \underline{\sigma}_{0,c}(x) =  \frac{1}{C} \sum_{b \in \Cc} \sigma_{0,c,b}(x),
\end{equation}
are coefficients that depend uniquely on the system geometry, frequency and pilot reuse, but are independent of the loading factor $S$ and of
the BS antenna factor $M$.
%Notice that in order to obtain (\ref{eq:bin-rate-lsubf}) we have used the fact that, by construction, the group spectral efficinecy
%is symmetric with respect to the cluster index $c$ and the frequency subband $f$.
\hfill \IEEEQED
\end{theorem}

%%%%%%%%%%%%%%%%%%%%%%%%%%%%%%%%%%%%%%%%%%
\section{Scheduling and Fairness}  \label{sec:scheduling}

Consider a system with $K$ bins,  $\{v(\Xc_0),\ldots, v(\Xc_{K-1})$, defined by sets $\Xc_k$ of symmetric locations chosen to uniformly discretize the
cellular coverage region $\Vc$.  The net bin spectral efficiency in bit/s/Hz, for each bin $v(\Xc_k)$, is obtained by
maximizing over all possible schemes in the family, i.e., over all possible clusters
$\Cc$ of size $C = 1,2,\ldots$,  frequency reuse factor $F$,  loading factor $S$, pilot reuse factor $Q$, and beamforming
scheme indicated by $J$, the product
\begin{equation} \label{ziocanissimo}
\max \{ 1 - QS/L, 0\}  \times R_{\Xc_k, \Cc}(F, C, J)
\end{equation}
where the first term takes into account
%
%denotes the ratio between data-phase and total slot channel uses,
%and takes into account
the pilot dimensionality overhead, and the second term is the spectral efficiency of the data phase
for a given network-MIMO scheme, given by Theorems \ref{J0thm}, \ref{Jgeq1C1thm} or \ref{Jgeq1Cgeq1thm}, depending on the case.
The maximization of (\ref{ziocanissimo}) is subject to the constraint $JS \leq CM$, which becomes relevant
for $J > 0$ (i.e., for LZFBF precoding).  Maximizing (\ref{ziocanissimo}) requires searching over a discrete parameter space
(apart from $S$, which is continuous). The simple closed-form expressions given in Theorems \ref{J0thm}, \ref{Jgeq1C1thm} and \ref{Jgeq1Cgeq1thm}
allow for an efficient system optimization, avoiding lengthy Monte Carlo simulations.

Suppose that for each bin $v(\Xc_k)$, the best scheme in the family of network-MIMO schemes
is found, and let $R^\star(\Xc_k)$ denote the corresponding maximum of (\ref{ziocanissimo}).
Then, a scheduler allocates the different bins on the time-frequency slots in order to maximize some
desired network utility function of the user rates. With randomized or round-robin selection of the active users in each bin,
each user in bin $v(\Xc_k)$ shares on average an equal fraction of the product  $\rho_k R^\star(\Xc_k)$, where $\rho_k$ is the fraction of time-frequency
slots allocated to bin $v(\Xc_k)$.
Under the assumption that users in the same bin should be treated with equal priority, we can focus on the maximization of
a componentwise non-decreasing concave network utility function of the bin spectral efficiencies, denoted by  $\Gc(R_0, \ldots, R_{K-1})$.
The scheduler determines the fractions $\{\rho_k\}$ by solving the following convex problem:
\begin{eqnarray} \label{eq:scheduling}
\mbox{maximize} & & \Gc(R_0, \ldots, R_{K-1}) \nonumber \\
\mbox{subject to} & & R_k \leq \rho_k R^\star(\Xc_k), \nonumber \\
& & \sum_{k=0}^{K-1} \rho_k \leq 1, \;\; \rho_k \geq 0.
\end{eqnarray}
For example, if Proportional Fairness (PF) \cite{Viswanath-Tse-Laroia-TIT02} is desired, we have
\begin{equation} \label{eq:PFS}
\Gc(R_0, \ldots, R_{K-1}) = \sum_{k=0}^{K-1} \log R_k,
\end{equation}
resulting in the bin time-frequency sharing fractions $\rho_k = 1/K$ (each bin is given an equal amount of slots).
In contrast, if the minimum user rate is relevant, we can impose max-min fairness by considering the function
\begin{equation} \label{eq:HFS}
\Gc(R_0, \ldots, R_{K-1}) = \min_{k = 0,\ldots, K-1} \; R_k.
\end{equation}
This results in the bin time-frequency sharing fractions $\rho_k = \frac{\frac{1}{R^\star(\Xc_k)}}{\sum_{j=0}^{K-1} \frac{1}{R^\star(\Xc_j)}}$.
More in general, a whole family of scheduling rules including (\ref{eq:PFS}) and (\ref{eq:HFS}) as special cases
is obtained by using the so-called $\alpha$-fairness network utility function, as defined in \cite{Mo-TNET00}.

%%%%%%%%%%%%%%%%%%%%%%%%%%%%%%%%%%%%%%%%%%%%%%%%%%%%%
\section{Numerical Results and Discussion} \label{sec:results}

In this section, we present some illustrative numerical results showing the following facts:
1) the asymptotic large-system analysis yields a very accurate approximation of the performance (obtained by monte Carlo simulation)
of actual finite-dimensional systems; 2) the proposed architecture based on partitioning the users' population in homogeneous bins and serving each bin with
specifically tailored network-MIMO scheme provides significant gains with respect to the ``Massive MIMO'' scheme of \cite{Marzetta-TWC10},
in the relevant regime of a finite number of BS antennas per active user.

At this point, it is worthwhile to make a comment on the convergence of finite-dimensional systems to the large-system limit as $N \rightarrow \infty$.
The approach of analyzing multiuser communication systems affected by random parameters (such as random channel matrices or random spreading matrices in CDMA)
in the limit of large dimension in order to exploit the rich, powerful and elegant theory of limiting distributions of large random matrices \cite{Tulino-Book04}
was pioneered  in \cite{Verdu-Shamai-TIT99,Tse-Verdu-TIT00} in the case of  random-spreading CDMA,
and successfully applied to single-user MIMO channels (see for example \cite{Chuah-Tse-TIT02,Lozano-Tulino-Verdu-TIT03,
Tulino-TIT05,Debbah-Muller-TIT05})
and to network-MIMO cellular systems \cite{Aktas-TIT06, Hoydis-Debbah-TSPsubmit, Zakhour-Hanly-JSACsubmit, Huh-Caire-TITsubmit, Huh-Tulino-Caire-TITsubmit}.
It was observed experimentally and proved mathematically (e.g., see \cite{Moustakas-TIT03, Kumar-Caire-TIT09}) that
the convergence of the actual finite-dimensional system spectral efficiency to the corresponding large-system limit  is very fast, as the system dimension $N$ increases.
In particular, well-known techniques can be used to analyze the ``fluctuation'' of the quantities of interest around their large-system limit for
large but finite $N$. Typically,  finite-$N$ ``concentration'' results are analogous to the Central Limit Theorem for i.i.d. random variables, but the convergence is much faster
owing to the fact that the eigenvalues of the matrices appearing in the spectral efficiency expressions are strongly correlated (see for example the discussion of
the results in \cite{Kumar-Caire-TIT09}). Since this convergence analysis is standard but cumbersome, and invariably points out that the large-system results are very good
predictions of the actual performance in cases of practical interest, here we focused only on the limit for $N \rightarrow \infty$
and provided a comparison with finite-dimensional simulation in order to corroborate our claims.

\begin{figure}
  \centering
  \includegraphics[width=10cm]{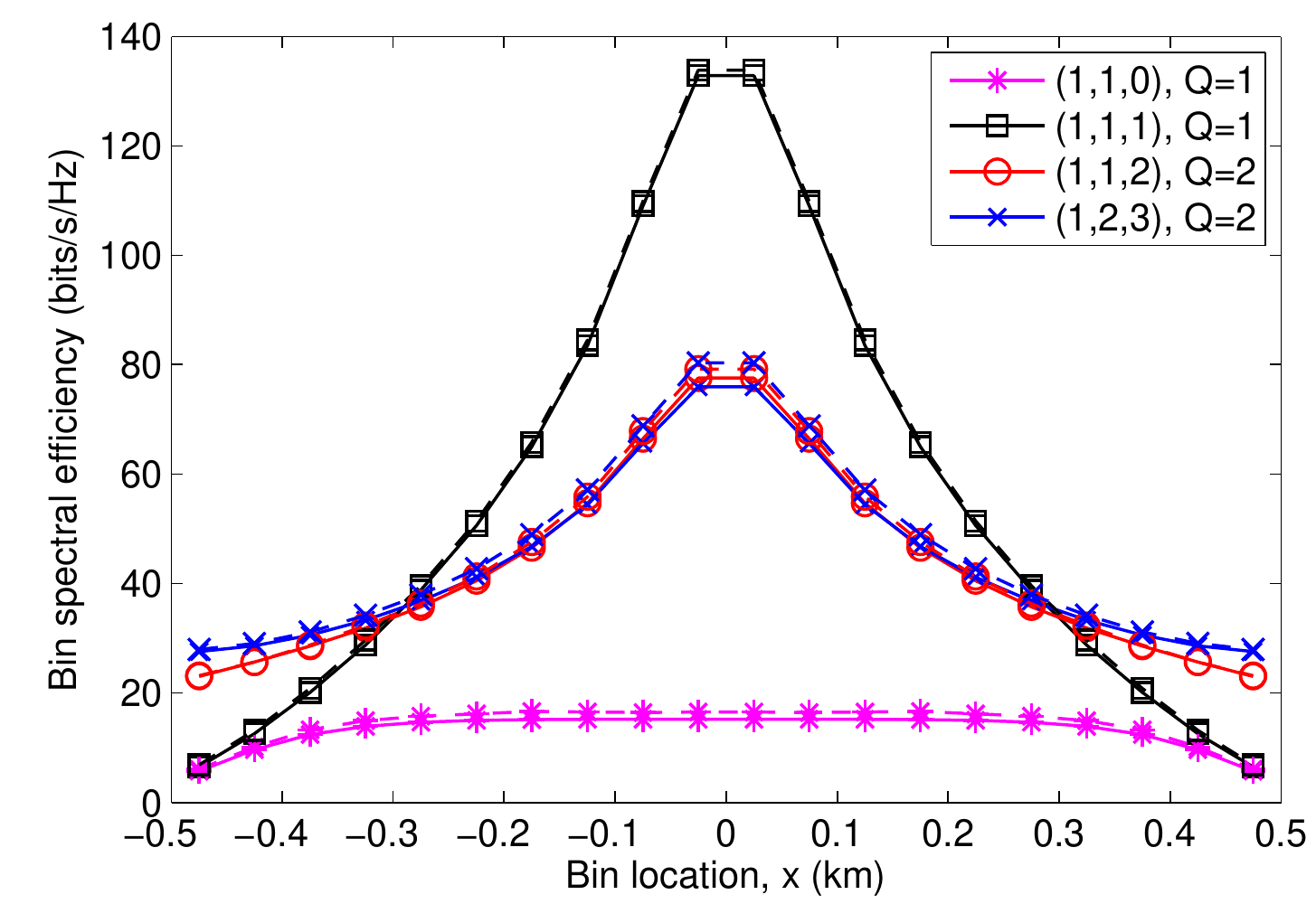}
  \caption{Bin spectral efficiency vs. location within a cell obtained from the large system analysis
  (solid) and the finite dimension ($N=1$) simulation (dotted) for various $(F,C,J)$.
  $M=30$ and $L=40$.}
  \label{fig:bin-rate-1dim}
\end{figure}

Fig.~\ref{fig:bin-rate-1dim} shows the group spectral  efficiency in (\ref{ziocanissimo}) as a function of the
bin locations within a cell  for different schemes identified by the parameters $(F,C,J)$ and $Q$.
The group spectral efficiency is obtained by Monte Carlo simulation (dotted) and
and is compared against the corresponding values from the closed-form
large-system analysis (solid), for the 1-dimensional cell layout of Fig.~\ref{fig:cell-cluster-bin}  with $B=24$ BSs,
$M=30$ antenna factor per BS,
$L=40$ coherence block dimension factor,
and $K=10$ bins in each cluster,
where clusters and location bins are given in Example~\ref{cluster-example} and \ref{binX-example},
with $x$ uniformly distributed in $[0,1/2]$. The pathloss model is the same as in \cite{Papadopoulos-Caire-Asilomar10}, where
$g(x,b) = G_0/(1 + (d_{\Lambda}(x,b)/\delta)^\alpha)$, with $G_0 = 10^6$, $\alpha = 3.76$,
and $\delta = 0.05$, and reflects (after suitable normalizations) a typical
cellular scenario with 1km diameter cells in a sub-urban environment.
The (1,1,1) scheme with $Q=1$ yields the best performance for locations near the cell center. However,
at the cell edges, $C=2$, $J=2$, or $F=2$ (not included in the figure)  attains significantly better performance.
As anticipated above, the limit for $N \rightarrow \infty$ matches very accurately with the Monte Carlo simuation even for very small
$N$ (we used the minimum possible $N=1$ in this case).
For this reason, in the following we present only the results for the large-system limit, obtained using the closed-form expressions of
Theorems \ref{J0thm}, \ref{Jgeq1C1thm} and \ref{Jgeq1Cgeq1thm}.

\begin{figure}
  \centering
  \subfigure[M=20]{
  \includegraphics[width=10cm]{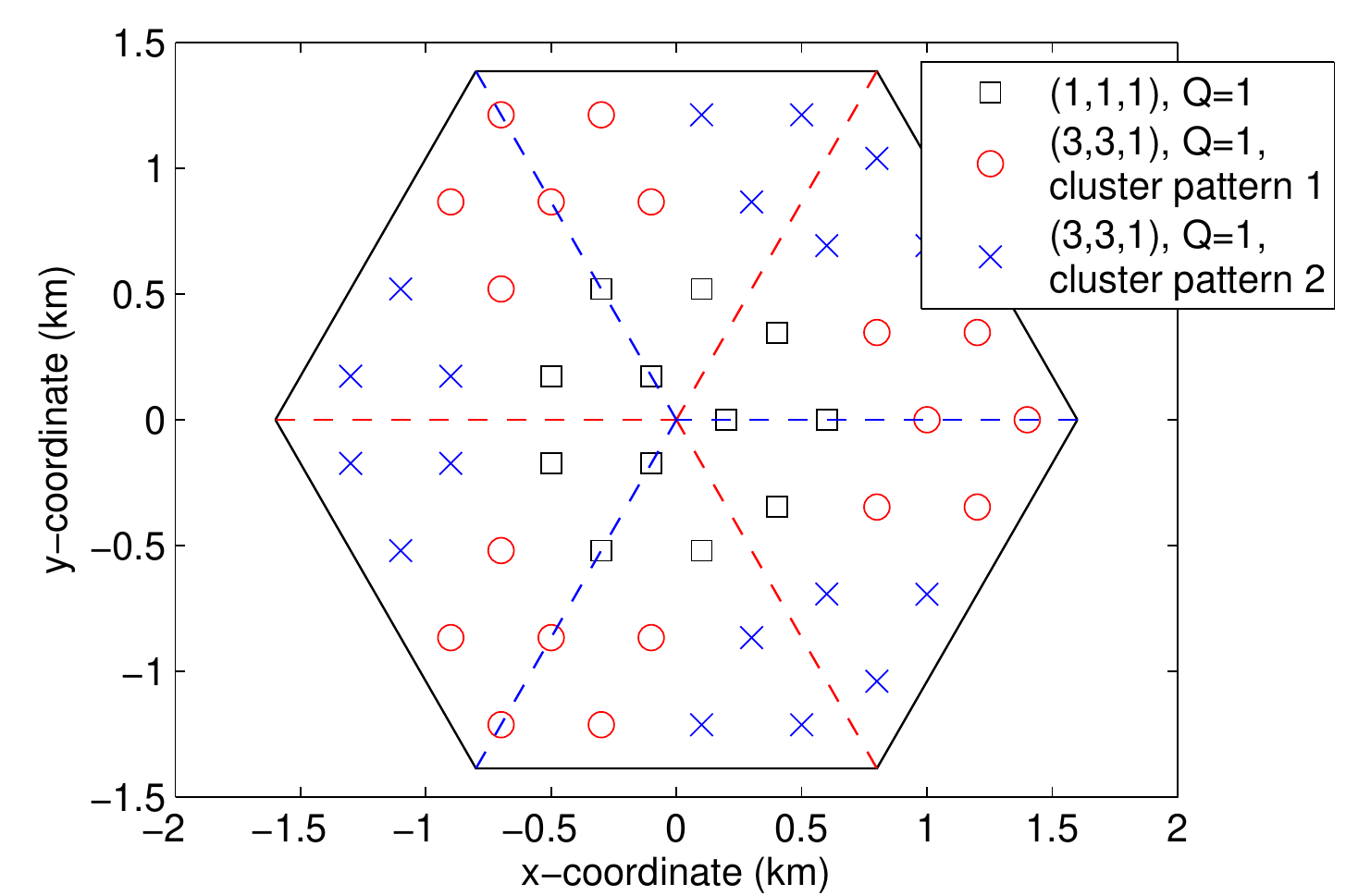}
  \label{subfig:opt-scheme-M20}}
  \subfigure[M=100]{
  \includegraphics[width=10cm]{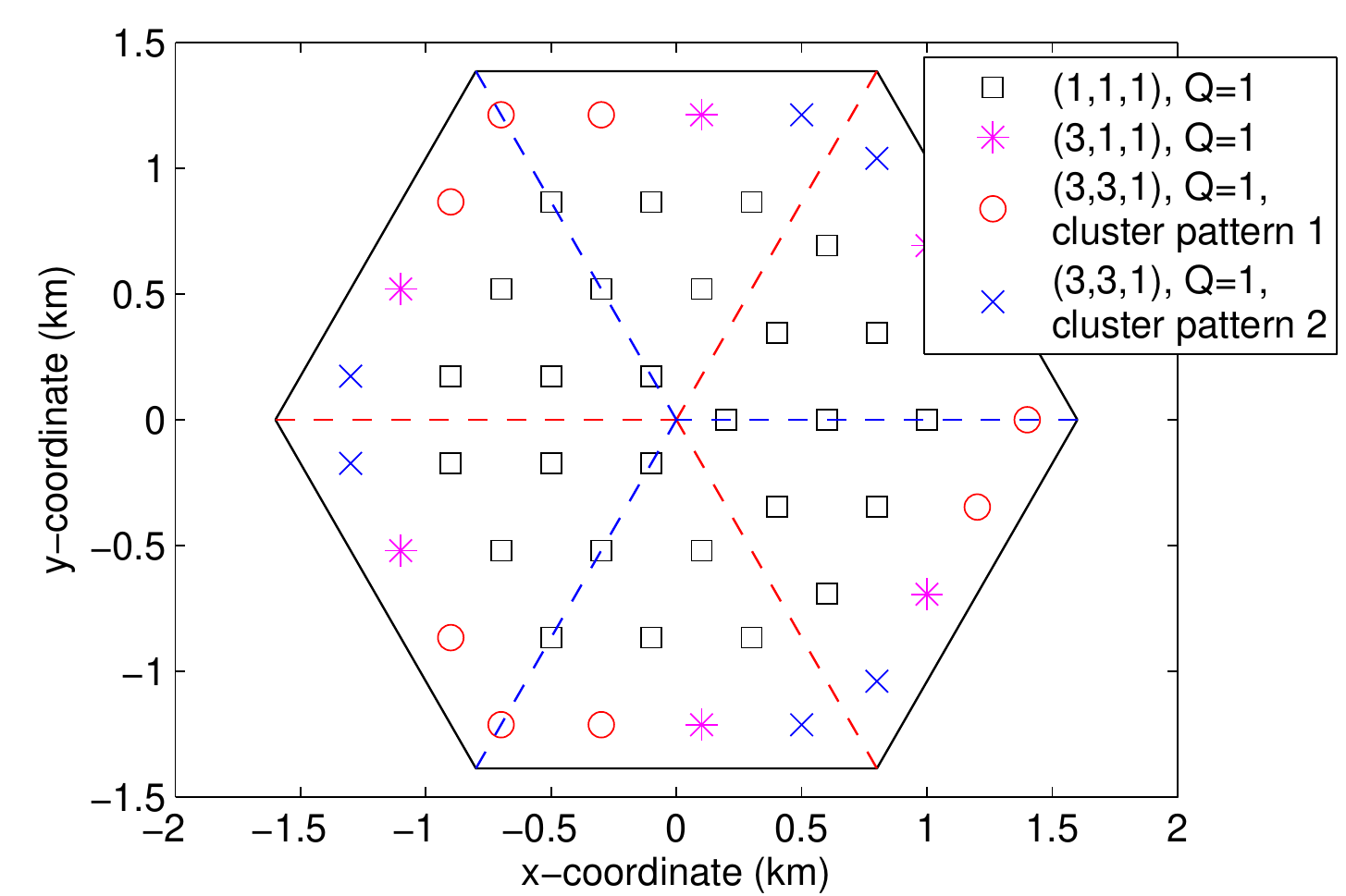}
  \label{subfig:opt-scheme-M100}}
  \caption{Optimal scheme at each user locations. $M=20$ and $100$,
  $K=16$, and $L=84$.}
  \label{fig:opt-scheme}
\end{figure}

In the 2-dimensional case, we considered the layout with $B=19$ hexagonal cells
as shown in Fig.~\ref{fig:2dim-lattice}. For comparison, we assume the same system model as in
\cite{Marzetta-TWC10}, with channel coherence block dimension,
the cell radius, and pathloss model given by $\Ltr = 84$,
1.6 km,  and $g(x,b)$ in the same form as before, with parameters $G_0 = 10^6$, $\delta = 0.1$ km, and $\alpha=3.8$,
respectively.
%, where $G_0$, the product of the transmit
%power and reference pathloss at distance $\delta$ is irrelevant under the settings of
%\cite{Marzetta-TWC10}.
Log-normal shadowing, considered in \cite{Marzetta-TWC10}, is not considered  here (see the comment in Section \ref{sec:conclusions}).
We considered schemes with cluster size $C=1$ and $C=3$, $K=16$ bins with 48 user locations,
where the cluster  and bin layout are qualitatively described in Fig.~\ref{fig:cell-cluster-bin}.
The frequency reuse factor $F$ and pilot reuse factor $Q$ are selected between
1 and 3 and, when $F$ or $Q=3$, the frequency subbands or training codebooks are allocated to
clusters as  shown in Fig.~\ref{fig:2dim-lattice} where different colors denotes different subbands or
training codebooks. Fig.~\ref{fig:opt-scheme} illustrates the optimum over the family of network-MIMO schemes
for (a) $M=20$ and (b) $M=100$. In both cases, $(1,1,1)$ is optimal in the inner part of the cell,
but schemes with $(3,3,1)$ or $(3,1,1)$ yield better performance for locations near cell boundary.
We notice also that the inner area within which the (1,1,1) scheme is
the best  increases with the BS antenna factor $M$.
%This suggest that when the number of BS antennas per active user is very large (towards the Massive MIMO limit),
%single-cell processing tends to overcome multi-cell joint processing, with the possible exception of
%a region near the cell edges, which shrinks as the number of antennas per active user increases.
%In contrast, when for no-to-large antennas per active user, multi-cell joint processing can achieve significant gains.

\begin{figure}
  \centering
  \includegraphics[width=10cm]{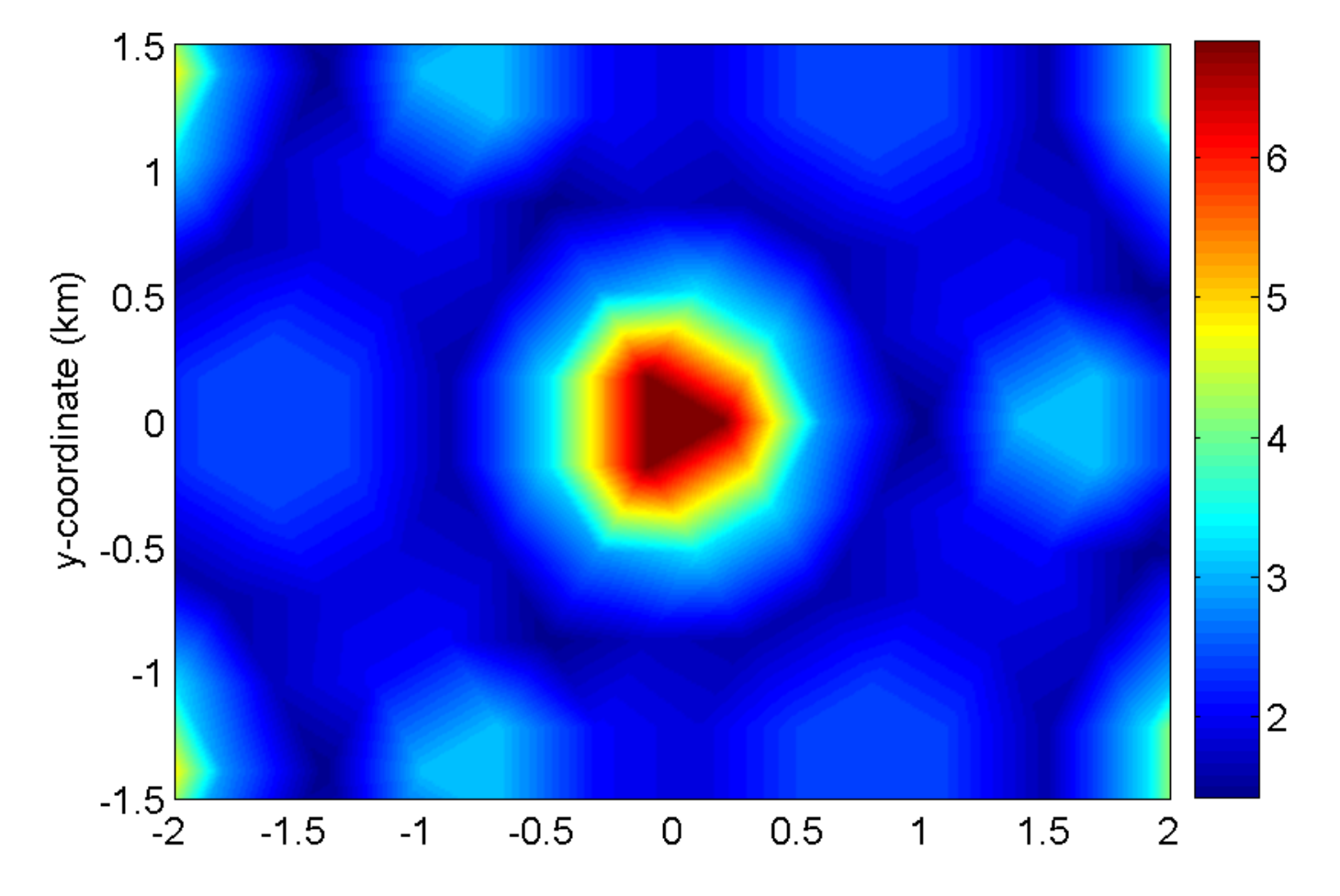}
  \caption{Bin-optimized spectral efficiencies normalized by the (1,1,0) spectral efficiencies.
  $M=50$, $K=48$,and $L=84$.}
  \label{fig:optimal-vs-mrt}
\end{figure}

\begin{figure}
  \centering
  \includegraphics[width=10cm]{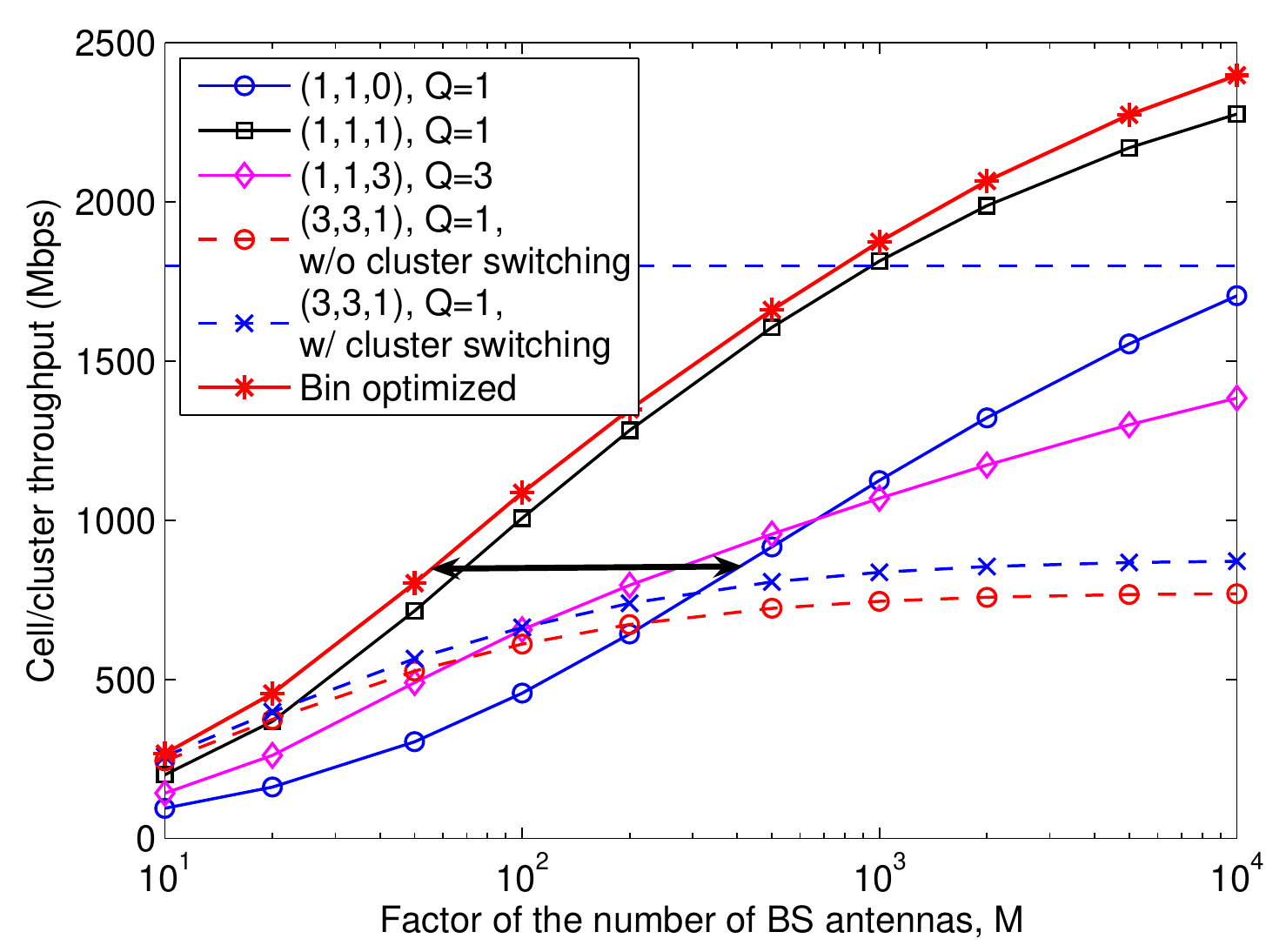}
  \caption{Cluster sum throughput vs. $M$ for various $(F,C,J)$
  and for a bin-optimized architecture under PF scheduling. $K=48$
  and $L=84$. The arrow indicates that the proposed architecture achieves the same spectral efficiency as the fixed
  scheme $(1,1,0)$ of \cite{Marzetta-TWC10}, with a 10-fold reduction of the number of BS antennas.}
  \label{fig:sum-throughput}
\end{figure}

Next, we compare the performance of the proposed architecture with the one advocated in
\cite{Marzetta-TWC10}. Fig.~\ref{fig:optimal-vs-mrt} shows the bin-optimized spectral
efficiency normalized by the  spectral efficiency of $(1,1,0),Q=1$ scheme (corresponding to \cite{Marzetta-TWC10}),
under two-dimensional layout with $M=50$.
The gain of the proposed architecture ranges from about 40\% to 580\%, depending on the users' location.
Fig.~\ref{fig:sum-throughput} shows the system throughput
as a function of $M$ in the two-dimensional layout. The throughput obtained for fixed parameters in the
considered family of network-MIMO schemes, as well as for the bin-optimized mixed-mode letting the scheduler choose the bin and
the associate network-MIMO scheme as described in Section \ref{sec:scheduling} is shown,
and compared with the reference performance of the $(1,1,0),Q=1$ scheme.
The cluster scheme includes two cases where the cluster pattern is fixed as one of two shown in
Fig.~\ref{fig:cell-cluster-bin} or can be switched to the closest one depending on the
user locations.  The system throughput of Fig.~\ref{fig:sum-throughput} is obtained  under PF scheduling (see (\ref{eq:PFS})).
For the sake of comparison, we assumed 20 MHz bandwidth and the coherence block size
$L=84$ as in \cite{Marzetta-TWC10} (considering the parameters of 3GPP LTE TDD system).
As the figure reveals, the $(3,3,1)$ schemes perform very well for small $M<20$ while, as $M$ increases, the $(1,1,1)$ scheme is best.
The bin-optimized architecture improves the throughput further at any value of $M$.
The dotted horizontal line in Fig.~\ref{fig:sum-throughput} denotes the cell throughput claimed
in \cite{Marzetta-TWC10} in the limit of an infinite number of transmit antennas per user with the
$(1,1,0), Q=1$ scheme.  We notice that this limit can be approached very slowly,   and more than 10000 antennas per BS are required (clearly impractical).
For finite number of antennas, the proposed architecture achieves the same throughput of
the scheme in \cite{Marzetta-TWC10} with a 10-fold reduction in the number of antennas at the base stations
(roughly, from 500 to 50 antennas,  as indicated by the arrow).

%%%%%%%%%%%%%%%%%%%%%%%%%%%%%%%%%%%%%%%%%%%%%%%%%%%%%
\section{Conclusions} \label{sec:conclusions}

We studied a novel network-MIMO TDD architecture that achieves
spectral efficiencies comparable with the recently proposed  ``Massive MIMO'' scheme,
with one order of magnitude less antennas per active user per cell.
The proposed strategy operates by partitioning the users population into geographically determined
``bins''. The time-frequency scheduling slots are allocated to the bins in order to form independent MU-MIMO
transmissions, each of which is optimized for the corresponding bin. This strategy allows the uplink training reuse factor,
the frequency reuse factor, the active  user loading factor, the BS cooperative cluster size and
the type of MU-MIMO linear beamforming to be finely tailored to the particular user bin.
We considered system optimization over 1-dimensional and 2-dimensional cell layouts,
based on a family of network-MIMO schemes ranging from
single-cell processing to joint processing over clusters of coordinated BSs, with linear precoders ranging from conventional
linear single-user beamforming
to zero-forcing beamforming with additional zero-forcing constraints for neighboring cells.
In order to carry out the system optimization, we developed efficient closed-form expressions for the achievable spectral efficiency for each scheme
in the family and each bin in the cellular layout. Our closed-form analysis is based on the large-system limit,
where all system dimensions scale to infinity with fixed ratios,  and make use of recent results
(by some of the authors of this paper) on the analysis of cellular systems with linear zero-forcing beamforming and channel
estimation errors \cite{Huh-Tulino-Caire-TITsubmit}. The performance predicted by the large-system asymptotic analysis is shown
to match very well with finite-dimensional simulations.
Our numerical results show that different schemes in the considered family achieve the best spectral
efficiency at different user locations.  This suggests the need for a location-adaptive scheme selection to serve efficiently
the whole coverage region. The resulting overall system is therefore a ``mixed-mode'' network-MIMO architecture, where different schemes, each of which is optimized for
the corresponding user bin, are multiplexed in the time-frequency plane.

As a final remark, it is worthwhile to point out that the approach of partitioning the users in homogeneous sets,
serving each set according to  a specifically optimized scheme,
and using  a scheduler to multiple different schemes in order to maximize some desired network utility function,
can be generalized to the case of shadowing, and to the case of users with different mobility.
This generalization is, however, non-trivial.  For example, in the presence of slow frequency-flat shadowing, ``bins''
are no-longer uniquely determined by the users geographic position. Rather, the set of large-scale channel gains (including shadowing), should be used
to classify the users in equivalence classes.
Also, in the presence of users with different mobility, users should be classified also on the basis of their different channel coherence block length.
The issue of how to optimally cluster users into equivalence classes that can be efficiently served in parallel, by MU-MIMO spatial multiplexing,
represents an interesting and important problem for future work.

\appendices

%%%%%%%%%%%%%%%%%%%%%%%%%%%%%%%%%%%%%%%%%%%%%%%%%%%%%
\section{Proof of Theorem \ref{J0thm}} \label{appen:bin-rate-mrt}

We focus on the reference cluster $\Cc$ (i.e., $c = 0$), with corresponding served group of locations
$\Xc = \{x_0,\ldots, x_{m-1}\}$.
For the sake of notation simplicity, we omit the subchannel index $f$,
and let $\Dc$ denote the set of clusters active on the same subchannel of
cluster 0, and $\Pc$ denote the set of clusters that share the same pilot block as cluster 0.
From (\ref{eq:received-signal}), the (scalar) signal received at some symbol interval of
the data phase, at the $k$-th active user receiver at location $x \in \Xc$, is given by
\begin{subequations} \label{eq:reciv-mrt}
\begin{eqnarray}
y_{k,0}(x) & = & u_{k,0}(x) \vv_{k,0}^\herm(x) \underline{\hv}_{k,0,0}(x) \label{useful-signal-term} \\
& & + \sum_{j\neq k} u_{j,0}(x) \vv_{j,0}^\herm(x) \underline{\hv}_{k,0,0}(x) +  \sum_{x' \in \Xc\backslash x} \sum_j u_{j,0}(x') \vv_{j,0}^\herm(x') \underline{\hv}_{k,0,0}(x)   \label{intra-cluster} \\
& & + \sum_{c' \in \Dc\backslash 0} \sum_{x' \in \Xc} \sum_j u_{j,c'}(x') \vv_{j,c'}^\herm(x') \underline{\hv}_{k,0,c'}(x) + z_{k,0}(x), \label{inter-cluster}
\end{eqnarray}
\end{subequations}
where $u_{j,c'}(x')$ denotes the code symbol
transmitted by cluster $c'$, to user $j$ at location $x' + c' : x'\in \Xc$.
With LSUBF downlink precoding, we have
\begin{equation} \label{lsubf-def}
\vv_{j,c'}(x') = \left \| \widehat{\underline{\hv}}_{j,c',c'}(x') \right \|^{-1}  \widehat{\underline{\hv}}_{j,c',c'}(x')
\end{equation}
Using the MMSE decomposition (\ref{eq:MMSE-decomp}), we isolate the {\em useful signal term}
from (\ref{useful-signal-term}), given by,
\begin{equation} \label{eq:useful-sig}
u_{k,0}(x) \vv_{k,0}^\herm(x) \underline{\widehat{\hv}}_{k,0,0}(x).
\end{equation}
The sum of the residual self-interference term due to the channel estimation error
with the signals in (\ref{intra-cluster}) transmitted by cluster 0 to the other users, results in the {\em intra-cluster interference term}
\begin{equation} \label{eq:intra-interf}
u_{k,0}(x) \vv_{k,0}^\herm(x) \underline{\ev}_{k,0,0}(x) + \sum_{j\neq k} u_{j,0}(x) \vv_{j,0}^\herm(x) \underline{\hv}_{k,0,0}(x) +
\sum_{x' \in \Xc\backslash x} \sum_j u_{j,0}(x') \vv_{j,0}^\herm(x') \underline{\hv}_{k,0,0}(x).
\end{equation}
Finally, the {\em ICI term} and background noise are given in (\ref{inter-cluster}).

A standard achievability bound based on the worst-case uncorrelated additive noise
\cite{Hassibi-Hochwald-TIT03} yields that the achievable rate
\begin{equation} \label{achievable-rate}
R_{k,0}^{(N)}(x) = \EE \left [ \log \left ( 1 + \frac{\EE\left [|\mbox{useful signal term}|^2 \; | \; \underline{\widehat{\hv}}_{k,0,0}(x) \right ]}
{\EE\left [ |\mbox{noise plus interference term}|^2 \; | \; \underline{\widehat{\hv}}_{k,0,0}(x) \right ]} \right ) \right ].
\end{equation}
%assuming that the receiver has perfect knowledge of its own estimated channel
%$\widehat{\underline{\hv}}_{k,0,0}(x)$.
%Next, we shall evaluate each conditional expectation in the large-system limit.
Both numerator and denominator of the
Signal-to-Interference plus Noise Ratio (SINR) appearing inside the log in (\ref{achievable-rate}) converge to deterministic limits as $N \rightarrow \infty$.
%Therefore, the outer expectation is not needed.
We will use extensively the representation of the channel MMSE estimates as
\begin{eqnarray} \label{representation}
\widehat{\underline{\hv}}_{j,c,c'}(x') & = & \frac{1}{\sqrt{N}} \Xim_{c,c'}^{1/2}(x') \av_{j,c,c'}(x')
\end{eqnarray}
where the vectors $\av_{j,c,c'}(x')$ are  i.i.d. $\sim \Cc\Nc(\zerov, \Id_{CMN})$, with generic components
denoted by $\{a_{n,b}: n = 1,\ldots, MN\}$ for all $b \in \Cc$.
We will also make use of the following limit, which follows as a direct application of the strong law of large numbers:
\begin{equation} \label{llnapplication}
\left \| \underline{\widehat{\hv}}_{j,c,c'}(x') \right \|^2 = \sum_{b\in \Cc} \xi_{c,c',b}(x') \frac{1}{N} \sum_{n=1}^{MN} |a_{n,b}|^2 \stackrel{{\rm a.s.}}{\rightarrow}
%CM \left ( \frac{1}{C} \sum_{b\in \Cc} \xi_{c,c',b}(x') \right ) \eqdef
CM \underline{\xi}_{c,c'}(x')
\end{equation}
where $\underline{\xi}_{c,c'}(x')$ is defined in (\ref{xibar}).  Using (\ref{lsubf-def}), the SINR numerator is given by
\begin{subequations} \label{eq:sinr-numer}
\begin{eqnarray}
\EE \left [ \left . \left | u_{k,0}(x)  \vv_{k,0}^\herm(x) \underline{\widehat{\hv}}_{k,0,0}(x) \right |^2 \right | \underline{\widehat{\hv}}_{k,0,0}(x) \right ]
& = &  \EE \left [ \left . \left | u_{k,0}(x) \right |^2 \left \| \underline{\widehat{\hv}}_{k,0,0}(x) \right \|^2 \right | \underline{\widehat{\hv}}_{k,0,0}(x) \right ] \label{use-lsubf} \\
& = & \frac{1}{S} \left \| \underline{\widehat{\hv}}_{k,0,0}(x) \right \|^2 \\
& \stackrel{{\rm a.s.}}{\rightarrow} & \frac{CM}{S} \underline{\xi}_{0,0}(x) \label{useful-signal-power}
\end{eqnarray}
\end{subequations}
where in (\ref{use-lsubf}) we used the LSUBF definition (\ref{lsubf-def}).

Next, we notice that all the terms forming interference and noise are uncorrelated. Hence, the conditional average interference power can be calculated
as a sum of individual terms. The self-interference due to non-ideal CSIT is given by
\begin{eqnarray} \label{self-interference}
\EE \left [ \left . \left | u_{k,0}(x) \vv_{k,0}^\herm(x) \underline{\ev}_{k,0,0}(x) \right |^2  \right | \underline{\widehat{\hv}}_{k,0,0}(x) \right ]
& = & \frac{1}{SN} \| \underline{\widehat{\hv}}_{k,0,0}(x) \|^{-2} \underline{\widehat{\hv}}^\herm_{k,0,0}(x) \Sigmam_{0,0}(x) \underline{\widehat{\hv}}_{k,0,0}(x) \nonumber \\
& = & \frac{1}{SN}
\frac{\sum_{b\in \Cc} \xi_{0,0,b}(x) \sigma_{0,0,b}(x) \frac{1}{N} \sum_{n=1}^{MN} |a_{n,b}|^2}{CM \underline{\xi}_{0,0}(x)} \nonumber \\
& \stackrel{{\rm a.s.}}{\rightarrow} & 0
\end{eqnarray}
where (\ref{self-interference}) follows from noticing that
\[
\sum_{b\in \Cc} \xi_{0,0,b}(x) \sigma_{0,0,b}(x) \frac{1}{N} \sum_{n=1}^{MN} |a_{n,b}|^2 \stackrel{{\rm a.s.}}{\rightarrow}  M \sum_{b\in \Cc} \xi_{0,0,b}(x) \sigma_{0,0,b}(x),  \]
which is a finite constant.

Following very similar calculations (omitted for brevity) and recalling that $g(x,b) = \xi_{0,0,b}(x) + \sigma_{0,0,b}(x)$
(see (\ref{eq:eff-pathloss})) and that
$SN/m$ users per location $x' \in \Xc$ are active,  we obtain
the intra-cluster interference power terms as
%\begin{eqnarray} \label{intra-interference1}
%\EE \left [ \left . \left | \sum_{j\neq k} u_{j,0}(x) \vv_{j,0}^\herm(x) \underline{\hv}_{k,0,0}(x) \right |^2 \right | \widehat{\underline{\hv}}_{k,0,0}(x) \right ]
%& \stackrel{{\rm a.s.}}{\rightarrow}  & \frac{1}{mC} \sum_{b \in \Cc} \frac{\xi_{0,0,b}(x) g(x,b)}{\underline{\xi}_{0,0}(x)}
%\end{eqnarray}
%and
%\begin{eqnarray} \label{intra-interference2}
%\EE \left [ \left . \left |   \sum_{x' \in \Xc\backslash x} \sum_j u_{j,0}(x') \vv_{j,0}^\herm(x') \underline{\hv}_{k,0,0}(x) \right |^2 \right | \widehat{\underline{\hv}}_{k,0,0}(x) \right ]
%& \stackrel{{\rm a.s.}}{\rightarrow}  & \frac{1}{mC} \sum_{x' \in \Xc\backslash x} \sum_{b \in \Cc} \frac{\xi_{0,0,b}(x') g(x,b)}{\underline{\xi}_{0,0}(x')}
%\end{eqnarray}
%so that, together, they yield the intra-cluster interference power term
\begin{equation} \label{intra-interference-tot}
\frac{1}{mC} \sum_{x' \in \Xc} \sum_{b \in \Cc} \frac{\xi_{0,0,b}(x') g(x,b)}{\underline{\xi}_{0,0}(x')}
\end{equation}
Next, we consider the ICI power term. In doing so, we must pay attention to the pilot contamination effect. In particular,
we have to separate all contributions in (\ref{inter-cluster}) coming from the $k$-th beam
of clusters $c' \in \Pc$ (i.e., for users sharing the same pilot signal of the reference user $k$ at $x \in \Xc$),
from the rest. The two contributions to the ICI are
\begin{equation} \label{same-pilot}
\Ic_{\mbox{same pilot}} = \sum_{c' \in \Pc \backslash 0} u_{k,c'}(x) \vv_{k,c'}^\herm(x) \underline{\hv}_{k,0,c'}(x)
\end{equation}
and
\begin{eqnarray} \label{no-same-pilot}
\Ic_{\mbox{no same pilot}} & = & \sum_{c' \in \Pc \backslash 0} \sum_{j\neq k} u_{j,c'}(x) \vv_{j,c'}^\herm(x) \underline{\hv}_{k,0,c'}(x)
+ \sum_{c' \in \Pc \backslash 0} \sum_{x' \in \Xc\backslash x} \sum_{j} u_{j,c'}(x') \vv_{j,c'}^\herm(x') \underline{\hv}_{k,0,c'}(x) \nonumber \\
& & + \sum_{c' \in \Dc - \Pc} \sum_{x' \in \Xc} \sum_j u_{j,c'}(x') \vv_{j,c'}^\herm(x') \underline{\hv}_{k,0,c'}(x)
\end{eqnarray}
Both  $\Ic_{\mbox{same pilot}}$ and $\Ic_{\mbox{no same pilot}}$ are independent of $\widehat{\underline{\hv}}_{k,0,0}(x)$.
Therefore, conditioning in the expectation can be omitted. Each individual term appearing in the sum (\ref{no-same-pilot}) yields
\begin{eqnarray*}
N \EE\left [ \left |u_{j,c'}(x') \vv_{j,c'}^\herm(x') \underline{\hv}_{k,0,c'}(x) \right |^2  \right ]
& \rightarrow & \frac{1}{S} \frac{\frac{1}{N} \trace \left ( \Xim_{c',c'}(x') \Gm_{0,c'}(x) \right )}{CM \underline{\xi}_{c',c'}(x')} \nonumber \\
& = & \frac{1}{SC} \sum_{b \in \Cc} \frac{\xi_{c',c',b}(x') g(x,c'+b)}{\underline{\xi}_{c',c'}(x')}.
\end{eqnarray*}
Summing over all terms, we have
\begin{equation} \label{no-same-pilot-ICI-power}
\EE \left [  |\Ic_{\mbox{no same pilot}}|^2  \right ] =
\sum_{c' \in \Dc\backslash 0} \frac{1}{mC} \sum_{x' \in \Xc} \sum_{b\in \Cc} \frac{\xi_{c',c',b}(x') g(x,c'+b)}{\underline{\xi}_{c',c'}(x')}.
\end{equation}
In order to evaluate $\EE \left [ |\Ic_{\mbox{same pilot}}|^2  \right ]$, we use
the decomposition (\ref{eq:MMSE-decomp1}) applied to $\underline{\hv}_{k,0,c'}(x)$, namely,
\begin{equation} \label{eq:MMSE-decomp2}
\underline{\hv}_{k,0,c'}(x) = \Gm_{0,c'}(x) \Gm_{c',c'}^{-1} (x)   \widehat{\underline{\hv}}_{k,c',c'}(x) + \underline{\ev}_{k,0,c'}(x).
\end{equation}
The general term in (\ref{same-pilot}) yields
\begin{align} \label{pilot-contamination-lsubf}
\EE & \left [  \left |u_{k,c'}(x) \vv_{k,c'}^\herm(x) \underline{\hv}_{k,0,c'}(x) \right |^2  \right ] \nonumber \\
& =\; \frac{1}{S} \EE \Big [ \| \widehat{\underline{\hv}}_{k,c',c'}(x) \|^{-2}
\left ( \widehat{\underline{\hv}}_{k,c',c'}^\herm (x) \Gm_{0,c'}(x) \Gm_{c',c'}^{-1} (x)   \widehat{\underline{\hv}}_{k,c',c'}(x) \right )^2  \nonumber \\
%&\quad + 2 \widehat{\underline{\hv}}_{k,c',c'}^\herm (x) \Gm_{0,c'}(x) \Gm_{c',c'}^{-1} (x)   \widehat{\underline{\hv}}_{k,c',c'}(x)  \Re \left \{
%\widehat{\underline{\hv}}^\herm_{k,c',c'}(x) \underline{\ev}_{k,0,c'}(x) \right \} \nonumber \\
&\quad + \widehat{\underline{\hv}}^\herm_{k,c',c'}(x) \underline{\ev}_{k,0,c'}(x)  \underline{\ev}^\herm_{k,0,c'}(x)  \widehat{\underline{\hv}}_{k,c',c'}(x) \Big ] \nonumber \\
& \rightarrow \; \frac{M}{SC \underline{\xi}_{c',c'}(x)} \left ( \sum_{b \in \Cc} \frac{g(x,c'+b)}{g(x,b)} \xi_{c',c',b}(x) \right )^2
\end{align}
where, inside the expectation,  we used the a.s. limits (\ref{llnapplication}),
\begin{eqnarray*}
\widehat{\underline{\hv}}^\herm_{k,c',c'}(x) \Gm_{0,c'}(x) \Gm_{c',c'}^{-1} (x)   \widehat{\underline{\hv}}_{k,c',c'}(x) & = &
\sum_{b\in \Cc} \frac{g(x,c'+b)}{g(c'+x,c'+b)} \xi_{c',c',b}(x) \frac{1}{N} \sum_{n=1}^{MN} |a_{n,b}|^2 \\
& \stackrel{{\rm a.s.}}{\rightarrow}  & M \sum_{b\in \Cc} \frac{g(x,c'+b)}{g(x,b)} \xi_{c',c',b}(x),
\end{eqnarray*}
with $g(c'+x,c'+b) = g(x,b)$, and $\widehat{\underline{\hv}}_{k,c',c'}^\herm(x) \underline{\ev}_{k,0,c'}(x) \stackrel{a.s.}{\rightarrow} 0$,  and the limit
\begin{eqnarray} \label{heeh}
\EE \left [ \widehat{\underline{\hv}}^\herm_{k,c',c'}(x) \underline{\ev}_{k,0,c'}(x)  \underline{\ev}^\herm_{k,0,c'}(x)  \widehat{\underline{\hv}}_{k,c',c'}(x) \right ]
& = & \frac{1}{N} \EE \left [ \widehat{\underline{\hv}}^\herm_{k,c',c'}(x) \Sigmam_{0,c'}(x) \widehat{\underline{\hv}}_{k,c',c'}(x) \right ]  \nonumber \\
& = & \frac{1}{N^2} \trace\left ( \Xim_{c',c'}(x) \Sigmam_{0,c'}(x) \right ) \rightarrow 0
\end{eqnarray}
Summing over all such terms, we obtain
\begin{equation} \label{same-pilot-ICI-power}
\EE \left [  |\Ic_{\mbox{same pilot}}|^2 \right ] =
\frac{CM}{S}  \sum_{c' \in \Pc \backslash 0}  \frac{1}{\underline{\xi}_{c',c'}(x)} \left ( \frac{1}{C} \sum_{b \in \Cc} \frac{g(x,c'+b)}{g(x,b)} \xi_{c',c',b}(x) \right )^2
\end{equation}
Using (\ref{useful-signal-power}), (\ref{intra-interference-tot}), (\ref{no-same-pilot-ICI-power}) and (\ref{same-pilot-ICI-power}) in (\ref{achievable-rate}),
recalling that the noise variance is equal to $1/F$, summing over all users in the reference group $\Xc$ and observing that the system is symmetric (by construction)
with respect to any cluster and any subband, we find the normalized group spectral efficiency of bin $v(\Xc)$ in the form (\ref{eq:bin-rate-lsubf}).

%%%%%%%%%%%%%%%%%%%%%%%%%%%%%%%%%%%%%%%%%%%%%%%%%%%%%
\section{Proof of Theorems \ref{Jgeq1C1thm} and \ref{Jgeq1Cgeq1thm}} \label{appen:bin-rate-zfbf}

With reference to Section \ref{precoding-types}, we consider LZFBF with or without inter-cluster interference (ICI) constraints, depending on the value of
$J \geq 1$. In particular, each cluster creates $JSN$ beamforming vectors for
$SN$ users in the same cluster and $(J-1)SN$ users in the neighboring clusters.
Any active user in the system, at any given scheduling slot,  imposes ZF constraints (see (\ref{ZFconstr}))
to $J$ clusters.  We restrict our attention to the three cases treated in Section \ref{precoding-types}, again referred to as
cases (a) $J=1$;
(b) $J=Q$; and
(c) $J=(Q-1)C+1$.
As before, we focus on the reference cluster $\Cc$ ($c = 0$) with served group $\Xc$.
In such cases, the beamforming matrix $\Vm_{0}$ is given by
the column-normalized Moore-Penrose pseudo-inverse of the estimated channel matrix (see (\ref{zfclassic}), (\ref{ziofa-caseb}) and (\ref{ziofa-casec})),
of size $CMN \times JSN$. This matrix is formed by blocks of size $N \times SN/m$, in the form:
\begin{equation} \label{block-matrix}
\Mm_0 = \left [ \begin{array}{cccc}
\Mm_{0,0} & \Mm_{0,1} & \cdots & \Mm_{0,Jm-1} \\
\Mm_{1,0} & \Mm_{1,1} & \cdots & \Mm_{1,Jm-1} \\
\vdots  &     &   & \vdots \\
\Mm_{CM-1,0} & \Mm_{CM-1,1} & \cdots  & \Mm_{CM-1, Jm-1} \end{array} \right ],
\end{equation}
where each block $\Mm_{i,j}$ corresponds to the $MN$ antennas of BS $b \in \Cc$ and to the $SN/m$ active users
in some location $x'$ with respect to which ZF constraints are imposed.
For the purpose of analysis, it is important to notice that the blocks  $\Mm_{i,j}$ are mutually independent,
and each block contains i.i.d. elements with mean zero and variance that depends on the block.
For example, if block $\Mm_{i,j}$ corresponds to a user location $c' + x': x' \in \Xc$ and BS $b \in \Cc$ such that the corresponding
channel vectors are estimated from the uplink training phase,  the elements of $\Mm_{i,j}$ are  $\sim \Cc\Nc(0, \xi_{c',0,b}(x)/N)$ (see (\ref{eq:eff-pathloss})
in Section \ref{subsec:mmse}). Instead, if block $\Mm_{i,j}$ corresponds to a user location and a BS such that the corresponding channel
vectors are treated as zero (see Section \ref{precoding-types}, Example \ref{JequalCQ-1+1}), then $\Mm_{i,j} = \zerov$ (all-zero block).

The signal received by user $k$ at location $x\in \Xc$ takes on the form (\ref{eq:reciv-mrt}).
From \cite[Theorem 3]{Huh-Tulino-Caire-TITsubmit} the rate
\begin{equation} \label{eq:achi-rate-lzfbf}
R_{k,0}^{(N)}(x) = \EE \left [ \log \left ( 1 + \frac{\EE\left [|\mbox{useful signal term}|^2 \; | \; \vv_{k,0}(x), \widehat{\underline{\hv}}_{k,0,0}(x) \right ]}
{\EE\left [ |\mbox{noise plus interference term}|^2 \; | \; \vv_{k,0}(x), \widehat{\underline{\hv}}_{k,0,0}(x)  \right]} \right ) \right ]
\end{equation}
is achievable, assuming that the receiver has perfect knowledge of its own estimated channel and beamforming vector.
The large-system limit  of the LZFBF useful signal coefficient $\vv_{k,0}^\herm(x) \underline{\widehat{\hv}}_{k,0,0}(x)$
for channel matrices in the form (\ref{block-matrix}) was obtained in \cite[Theorem 1]{Huh-Tulino-Caire-TITsubmit}
(details are omitted for the sake of brevity).
While in general this limit is obtained as the solution of a fixed-point equation that must be solved numerically,
the user locations and the BS positions considered in this paper satisfy the symmetry conditions given
in \cite[Section III.A]{Huh-Tulino-Caire-TITsubmit}, and the asymptotic useful signal term admits  a simple closed
form given in \cite[eq. (32)]{Huh-Tulino-Caire-TITsubmit}. Applying this result we obtain
\begin{eqnarray} \label{useful-signal-zf-result}
\left | \vv_{j,c'}^\herm(x') \underline{\widehat{\hv}}_{j,c',c'}(x') \right |^2 & \stackrel{{\rm a.s.}}{\rightarrow} &
(CM - JS) \underline{\xi}_{c',c'}(x')
\end{eqnarray}
for any $j, c'  \Lambda_{\rm bs} \cap \Vc$ and $x' \in \Xc$.
By construction, it is assumed that $JS < CM$.
Notice the well-known dimensionality limit of the ZF beamforming: when the ratio of the number of ZF constraints per degree of freedom (antenna)
$JS/(CM)$ tends to 1, the effective useful signal term vanishes.
Using (\ref{useful-signal-zf-result}) and recalling that $\EE[|u_{k,0}(x)|^2] = 1/S$
we obtain the SINR numerator in (\ref{eq:achi-rate-lzfbf}) as
\begin{equation} \label{eq:useful-sig-pwr-lzfbf}
\EE \left[ \left. \left| u_{k,0}(x) \vv_{k,0}^\herm(x) \underline{\widehat{\hv}}_{k,0,0}(x) \right|^2
\right | \vv_{k,0}(x), \widehat{\underline{\hv}}_{k,0,0}(x) \right]
\stackrel{{\rm a.s.}}{\rightarrow}  \frac{CM-JS}{S} \underline{\xi}_{0,0}(x).
\end{equation}
As done in Appendix \ref{appen:bin-rate-mrt}, we consider the intra-cluster, ICI and noise terms in the
SINR denominator of (\ref{eq:achi-rate-lzfbf}) separately.
The ZF constraints imply that $\vv_{j,0}^\herm(x') \underline{\widehat{\hv}}_{k,0,0}(x) = 0$ for all
$(j,x') \neq (k,x) : x' \in \Xc$. Therefore,  the intra-cluster interference term given in general by
(\ref{eq:intra-interf}), reduces to
\[ \sum_{x' \in \Xc} \sum_j u_{j,0}(x') \vv_{j,0}^\herm(x') \underline{\ev}_{k,0,0}(x), \]
and its conditional second moment is given by
\begin{align} \label{eq:intra-intf-lzfbf}
& \EE \left[ \left. \left| \textstyle{\sum_{x' \in \Xc} \sum_j} u_{j,0}(x') \vv_{j,0}^\herm(x')
  \underline{\ev}_{k,0,0}(x) \right|^2 \right | \vv_{k,0}(x), \widehat{\underline{\hv}}_{k,0,0}(x)  \right] \nonumber \\
& = \;
  \frac{1}{S} \EE \left [ \left . \trace \left( \Vm_0^\herm \EE \left[ \underline{\ev}_{k,0,0}(x)
  \underline{\ev}^\herm_{k,0,0}(x) \right] \Vm_0 \right) \right | \vv_{k,0}(x) \right ] \nonumber \\
&= \;  \frac{1}{SN} \EE \left [ \left . \trace \left( \Vm_0 \Vm_0^\herm \Sigmam_{0,0}(x) \right) \right | \vv_{k,0}(x) \right ] \nonumber \\
&\stackrel{{\rm a.s.}}{\rightarrow} \;  \frac{1}{C} \sum_{b \in \Cc} \sigma_{0,0,b}(x)
  \eqdef \underline{\sigma}_{0,0}(x),
\end{align}
where the last line holds from the following lemma\footnote{Notice that
since the columns of $\Vm_0$ have unit norm we have $\frac{1}{N} \trace(\Vm_0 \Vm_0^\herm)
= \frac{1}{N} \trace(\Vm_0^\herm \Vm_0) = S$. However, since $\Sigmam_{0,0}(x)$ is block-diagonal
with constant diagonal blocks $\sigma_{0,0,b}(x) \Id_{MN}$, the constant partial trace property is
needed in order to obtain (\ref{eq:intra-intf-lzfbf}).}.

\begin{lemma} \label{lem:constant-partial-trace}
If the user locations and BS positions are symmetric (in the sense defined in
 \cite[Section III.A]{Huh-Tulino-Caire-TITsubmit}, which is satisfied for the choice of lattice-based user locations sets considered in this work),
 the matrix $\Vm_c \Vm_c^\herm$ satisfies the
``constant partial trace'' property in the large-system limit, i.e.,
the sum of a block of $MN$ consecutive diagonal elements of $\Vm_c \Vm_c^\herm$ corresponding
to the antennas of BS $b \in \Cc$, divided by $SN$, tends to the constant limit $1/C$, independent of the BS index $b$.
\end{lemma}
\begin{IEEEproof}
The sum of the $b$-th diagonal element block of size $MN$ of $\Vm_c \Vm_c^\herm$,
divided by $SN$, is written as
\begin{align} \label{eq:per-BS-power}
\frac{1}{SN} \sum_{\ell=(b-1)MN+1}^{bMN} \left[ \Vm_c \Vm_c^\herm \right]_{\ell,\ell}
\;=\; \frac{1}{SN} \sum_{\ell=(b-1)MN+1}^{bMN} \sum_{x \in \Xc} \sum_{j=1}^{SN/m}
  \left| \left[\vv_{j,c} (x) \right]_{\ell} \right|^2,
\end{align}
where $[\vv_{j,c}(x)]_\ell$ denotes the $\ell$-th element of the column $\vv_{j,c}(x)$ of $\Vm_c$.
Next, for the sake of clarity,
we identify some terms in the notation of this paper with the corresponding
terms in the notation of \cite[Lemma 1]{Huh-Tulino-Caire-TITsubmit}.
To this purpose, we enumerate the locations $x \in \Xc$ as $ k = 1, \ldots, m$.
The transmit power to  each active user in the cluster is
$q_k = 1/S$ and  the fraction of active users per location is $\mu_k = S/m$. Both quantities are constant with $k$.
Also, the term
\[ \frac{1}{N} \sum_{\ell=(b-1)MN+1}^{bMN}  \sum_{j=1}^{SN/m} \left | \left[\vv_{j,c} (x) \right]_{\ell} \right|^2, \]
coincides with the term $\theta_{b,k}$ defined in \cite[eq. (26)]{Huh-Tulino-Caire-TITsubmit}
Therefore, the term in (\ref{eq:per-BS-power}) can be written as
$\frac{1}{S} \sum_{k=1}^m \theta_{b,k} = \sum_{k=1}^m q_k \theta_{b,k}$.
Applying \cite[Lemma 1]{Huh-Tulino-Caire-TITsubmit}, we have
$$\sum_{k=1}^m q_k \theta_{b,k} =  \sum_{k=1}^{m/C} q_k \mu_k = \frac{m}{C} \cdot \frac{1}{S} \cdot \frac{S}{m} = \frac{1}{C}.$$
\end{IEEEproof}

Next, we consider the ICI term and we separate it into $\Ic_{\mbox{same pilot}}$ and $\Ic_{\mbox{no same pilot}}$.
The conditioning with respect to $\vv_{k,0}(x), \widehat{\hv}_{k,0,0}(x)$ is irrelevant for the ICI terms and therefore it can be omitted.
First, we evaluate the pilot contamination effect for the case $C = 1$.
Using (\ref{eq:MMSE-decomp2}), (\ref{eq:useful-sig-pwr-lzfbf}) and (\ref{same-pilot}), we obtain
\begin{eqnarray} \label{eq:same-pilot-ICI-term-lzfbf-C1}
\EE \left[  |\Ic_{\mbox{same pilot}}|^2 \right] &= & \EE\left[ \left| \sum_{c' \in \Pc\backslash 0} u_{k,c'}(x) \vv_{k,c'}^\herm(x) \underline{\hv}_{k,0,c'}(x) \right|^2  \right] \nonumber \\
&= & \frac{1}{S} \sum_{c' \in \Pc \backslash 0} \EE \left[ \left| \vv^\herm_{k,c'}(x) \left( \Gm_{0,c'}(x) \Gm_{c',c'}^{-1}(x)
  \widehat{\underline{\hv}}_{k,c',c'}(x) + \underline{\ev}_{k,0,c'}(x)\right) \right|^2 \right] \nonumber \\
&= & \frac{1}{S} \sum_{c' \in \Pc\backslash 0} \left \{ \left ( \frac{g(x,c')}{g(x,0)} \right)^2 \EE \left[ \left| \vv^\herm_{k,c'}(x) \widehat{\underline{\hv}}_{k,c',c'}(x) \right|^2 \right]
  + \EE \left[ \left| \vv^\herm_{k,c'}(x) \underline{\ev}_{k,0,c'}(x) \right|^2 \right] \right \} \nonumber \\
& \rightarrow &
  \frac{M-JS}{S} \sum_{c' \in \Pc\backslash 0}  \left ( \frac{g(x,c')}{g(x,0)} \right)^2 \underline{\xi}_{c',c'}(x)
\end{eqnarray}
where we used (\ref{useful-signal-zf-result}) and
\begin{eqnarray} \label{veev}
\EE \left[ \left| \vv^\herm_{k,c'}(x) \underline{\ev}_{k,0,c'}(x) \right|^2 \right]
& = & \frac{1}{N} \EE \left [ \vv^\herm_{k,c'}(x)  \Sigmam_{0,c'}(x) \vv_{k,c'}(x) \right ] \nonumber \\
& \leq & \frac{1}{N} \EE \left [ \| \vv_{k,c'}(x) \|^2 \right ] \max_{b \in \Cc} \{ \sigma_{0,c',b}(x) \} \nonumber \\
& = & \frac{1}{N}  \max_{b \in \Cc} \{ \sigma_{0,c',b}(x)  \} \rightarrow 0
\end{eqnarray}
For $C > 1$, we have $\Gm_{0,c'}(x) \Gm_{c',c'}^{-1}(x) = \diag\left ( \frac{g(x, c'+b)}{g(x,b)} \Id_{MN} : b \in \Cc \right )$.
While $\vv^\herm_{k,c'}(x)$ and $\widehat{\underline{\hv}}_{k,c',c'}(x)$ are orthogonal by design,
the term $\EE \left [ \left | \vv^\herm_{k,c'}(x) \Gm_{0,c'}(x) \Gm_{c',c'}^{-1}(x) \widehat{\underline{\hv}}_{k,c',c'}(x) \right |^2 \right ]$ is generally
non-zero and does not admit a simple closed-form since $\vv_{k,c'}(x)$ and $\widehat{\underline{\hv}}_{k,c',c'}(x)$ are statistically dependent.
In order to overcome this problem, we consider the following upper bound  obtained by applying Cauchy-Schwartz inequality:
\begin{eqnarray} \label{cauchy-schwartz}
\left | \vv^\herm_{k,c'}(x) \Gm_{0,c'}(x) \Gm_{c',c'}^{-1}(x) \widehat{\underline{\hv}}_{k,c',c'}(x) \right |^2 & \leq &
\left \| \vv_{k,c'}(x) \right \|^2  \left \| \Gm_{0,c'}(x) \Gm_{c',c'}^{-1}(x) \widehat{\underline{\hv}}_{k,c',c'}(x) \right \|^2
\end{eqnarray}
Recalling that $\vv_{k,c'}(x)$ has unit norm and that $\EE [ \widehat{\underline{\hv}}_{k,c',c'}(x) \widehat{\underline{\hv}}^\herm_{k,c',c'}(x) ] = \frac{1}{N} \Xim_{c',c'}(x)$, we obtain
\begin{equation} \label{eq:same-pilot-ICI-term-lzfbf-Ccluster-bound}
\EE \left[  |\Ic_{\mbox{same pilot}}|^2 \right]
\leq  \frac{CM}{S} \sum_{c' \in \Pc\backslash 0}  \frac{1}{C} \sum_{b \in \Cc} \left ( \frac{g(x,c'+b)}{g(x,b)} \right)^2 \xi_{c',c',b}(x)
\end{equation}
%The approximation, widely validated by extensive comparison with finite-dimensional simulations,
%is obtained by replacing  $\Gm_{0,c'}(x) \Gm_{c',c'}^{-1}(x)$ with a scaled identity matrix with the same trace. This yields
%\begin{equation} \label{eq:same-pilot-ICI-term-lzfbf-Ccluster-approx}
%\EE \left[  |\Ic_{\mbox{same pilot}}|^2 \right]
%\approx  \frac{CM -JS}{S} \sum_{c' \in \Pc\backslash 0}  \left ( \frac{1}{C} \sum_{b \in \Cc} \frac{g(x,c'+b)}{g(x,b)} \right)^2 \underline{\xi}_{c',c'}(x)
%\end{equation}
%Notice that for $C = 1$ (\ref{eq:same-pilot-ICI-term-lzfbf-Ccluster-approx}) coincides with the exact expression
%(\ref{eq:same-pilot-ICI-term-lzfbf-C1}), while the bound (\ref{eq:same-pilot-ICI-term-lzfbf-Ccluster-bound}) is not tight even for $C = 1$.}

Next, we examine the ICI power caused by the term $\Ic_{\mbox{no same pilot}}$.
In the case of $J \geq 2$, this can be further decomposed into a term
$\Ic_{\mbox{ICI-ZF}}$, taking into account the clusters which have a ZF constraint with respect to user $k$ at location
$x \in \Xc$,  and $\Ic_{\mbox{no-ICI-ZF}}$, taking into account all other clusters.
In order to proceed, we define $\Ec(x)$ as the set of $J-1$
clusters $c \neq 0$ with centers closest to $x \in \Xc$.
With these definition, we have
\begin{equation} \label{eq:I-ici-const}
\Ic_{\mbox{ICI-ZF}} \;=\; \sum_{c' \in \Ec(x)} \sum_{x' \in \Xc} \sum_{j}
  u_{j,c'}(x') \vv_{j,c'}^\herm(x') \underline{\hv}_{k,0,c'}(x)
\end{equation}
and
\begin{align} \label{eq:I-no-ici-const}
\Ic_{\mbox{no-ICI-ZF}} \;=\;& \sum_{c' \in \Pc\backslash 0} \sum_{j\neq k} u_{j,c'}(x)
  \vv_{j,c'}^\herm(x) \underline{\hv}_{k,0,c'}(x)
  + \sum_{c' \in \Pc\backslash 0} \sum_{x' \in \Xc\backslash x} \sum_{j} u_{j,c'}(x')
  \vv_{j,c'}^\herm(x') \underline{\hv}_{k,0,c'}(x) \nonumber \\
&+ \sum_{c' \in \Dc - \Pc - \Ec(x)} \sum_{x' \in \Xc} \sum_j u_{j,c'}(x')
  \vv_{j,c'}^\herm(x') \underline{\hv}_{k,0,c'}(x).
\end{align}
We start with the terms in (\ref{eq:I-no-ici-const}).
For $c' \in \Pc\backslash 0$, by definition of LZFBF we have that
$\vv_{j,c'}^\herm(x') \widehat{\underline{\hv}}_{k,c',c'}(x) = 0$ for all
$(j,x') \neq (k,x)$.  For $C = 1$, since $\Gm_{0,c'}(x) \Gm_{c',c'}^{-1} (x)$ is a scaled identity matrix, using
(\ref{eq:MMSE-decomp2}) we have that
\begin{equation} \label{zzz}
\vv_{j,c'}^\herm(x') \underline{\hv}_{k,0,c'}(x) =  \vv_{j,c'}^\herm(x') \underline{\ev}_{k,0,c'}(x)
\end{equation}
For $c' \in \Dc - \Pc - \Ec(x)$, the vectors $\vv_{j,c'}^\herm(x')$ and  $\underline{\hv}_{k,0,c'}(x)$ are statistically independent.
Hence, for $C = 1$ we have
\begin{align} \label{eq:no-ici-const-ici-power-lzfbfC1}
& \lim_{N \rightarrow \infty} \EE \left [ \left | \Ic_{\mbox{no-ICI-ZF}}  \right |^2 \right ] \nonumber \\
& =  \; \lim_{N \rightarrow \infty} \left \{ \sum_{c' \in \Pc \backslash 0} \frac{1}{SN} \EE \left [ \trace \left ( \Vm_{c'} \Vm_{c'}^\herm \Sigmam_{0,c'}(x) \right ) \right ] +
\sum_{c' \in \Dc-\Pc-\Ec(x)} \frac{1}{SN} \EE \left [ \trace \left ( \Vm_{c'} \Vm_{c'}^\herm \Gm_{0,c'}(x) \right ) \right ] \right \} \nonumber \\
& = \; \sum_{c' \in \Pc\backslash 0} \sigma_{0,c',0}(x) + \sum_{c' \in \Dc - \Pc - \Ec(x)} g(x,c')
\end{align}
where in (\ref{eq:no-ici-const-ici-power-lzfbfC1}) we used
Lemma~\ref{lem:constant-partial-trace} for matrix $\Vm_{c'} \Vm_{c'}^\herm$.

For $C > 1$, because of the block-diagonal form of the matrix $\Gm_{0,c'}(x) \Gm_{c',c'}^{-1}(x)$ already mentioned before,
(\ref{zzz}) does not hold in general.  An upper bound to the interference power in this case
can be obtained by assuming that the MMSE estimate $\widehat{\underline{\hv}}_{k,0,c'}(x)$ of the channel from
user $k$ at location $x \in \Xc$ and the antennas of cluster $c'$ is so noisy that it can be considered equal to zero.
Therefore, the estimation error $\ev_{k,0,c'}(x)$ has covariance $\frac{1}{N} \Gm_{0,c'}(x)$, and we obtain
\begin{equation} \label{eq:no-ici-const-ici-power-lzfbfCclutser-bound}
\lim_{N \rightarrow \infty} \EE \left [ \left | \Ic_{\mbox{no-ICI-ZF}}  \right |^2 \right ]
\leq \sum_{c' \in \Dc\backslash 0 \; - \; \Ec(x)} \underline{g}_{0,c'}(x)
\end{equation}
where $\underline{g}_{0,c'}(x)$ is defined in (\ref{gg}).
Finally, we consider $\Ic_{\mbox{ICI-ZF}}$ in (\ref{eq:I-ici-const}). We distinguish different cases depending on the
value of $J$.  In case (a), this term is zero.
In case (b), we have $\vv_{j,c'}^\herm(x') \widehat{\underline{\hv}}_{k,0,c'}(x)=0$ for all $j$ and all $x \in \Ec(x)$. Hence, similarly to (\ref{eq:intra-intf-lzfbf}),
we obtain
\begin{equation} \label{eq:ici-const-pwr-b}
\EE \left[ \left|\Ic_{\mbox{ICI-ZF}}\right|^2  \right]
\; \rightarrow \; \sum_{c' \in \Ec(x)} \underline{\sigma}_{0,c'}(x)
\end{equation}
In case (c), the ZF vectors $\vv_{j,c'}(x')$ of cluster $c' \in \Ec(x)$ are calculated by imposing orthogonality conditions
with the segment of the estimated channel vector $\widehat{\underline{\hv}}_{k,0,c'}(x)$ corresponding to the $MN$ antennas of the closest BS.
In order to proceed further, we define the index of the closest BS to location $x$ in cluster $c' \in \Ec(x)$ as
$b(x,c) = \arg\min\{ d_{\Lambda}(x,c+b) : b \in \Cc\}$. Then, the effective channel used for ZF beamforming calculation is given by
\[ \underline{\widetilde{\hv}}_{k,0,c'}(x) = \Psim_{b(x,c')} \widehat{\underline{\hv}}_{k,0,c'}(x) \]
where  $\Psim_{b(x,c')}$ is a selection matrix, with all elements equal to zero but for a block of diagonal elements corresponding to
the positions of the $MN$ antennas of BS $b(x,c')$.  By construction, and using the MMSE decomposition, we have
\begin{eqnarray}
\vv_{j,c'}^\herm(x') \underline{\hv}_{k,0,c'}(x) & = & \vv_{j,c'}^\herm(x')
\left ( \Psim_{b(x,c')} \widehat{\underline{\hv}}_{k,0,c'}(x)  +
(\Id_{CMN} - \Psim_{b(x,c')} ) \widehat{\underline{\hv}}_{k,0,c'}(x)  + \ev_{k,0,c'}(x) \right ) \nonumber \\
& = & \vv_{j,c'}^\herm(x') \left ( (\Id_{CMN} - \Psim_{b(x,c')} ) \underline{\hv}_{k,0,c'}(x)  + \Psim_{b(x,c')} \ev_{k,0,c'}(x) \right ) \nonumber \\
& = & \vv_{j,c'}^\herm(x') \widetilde{\ev}_{k,0,c'}(x)
\end{eqnarray}
where $\widetilde{\ev}_{k,0,c'}(x)$ is independent of all beamfomrming vectors  $\Vm_{c'}$ of cluster $c' \in \Ec(x)$, and has covariance matrix
\[ \frac{1}{N} \left ( (\Id_{CMN} - \Psim_{b(x,c')} )  \Gm_{0,c'}(x) (\Id_{CMN} - \Psim_{b(x,c')} ) + \Psim_{b(x,c')} \Sigmam_{0,c'}(x) \Psim_{b(x,c')}  \right ) \]
Using these facts and operating similarly as in (\ref{eq:ici-const-pwr-b}), we obtain
\begin{equation} \label{eq:ici-const-pwr-c}
\EE \left[ \left|\Ic_{\mbox{ICI-ZF}}\right|^2  \right]
\; \rightarrow \; \sum_{c' \in \Ec(x)} \frac{1}{C} \left ( \sum_{b \in \Cc\backslash b(x,c')} g(x, b +c')  + \sigma_{0,c',b(x,c')}(x) \right )
\end{equation}
From (\ref{eq:useful-sig-pwr-lzfbf}), (\ref{eq:intra-intf-lzfbf}),
(\ref{eq:same-pilot-ICI-term-lzfbf-C1}),
%(\ref{eq:same-pilot-ICI-term-lzfbf-Ccluster-bound}),
(\ref{eq:no-ici-const-ici-power-lzfbfC1}),
%(\ref{eq:no-ici-const-ici-power-lzfbfCclutser-bound}),
(\ref{eq:ici-const-pwr-b}), and (\ref{eq:ici-const-pwr-c}),
the normalized group spectral efficiency for $C = 1$ and $J \geq 1$ is obtained in the form (\ref{eq:bin-rate-zfbf})
For the cluster case $C > 1$, using bounds (\ref{eq:same-pilot-ICI-term-lzfbf-Ccluster-bound}), and
(\ref{eq:no-ici-const-ici-power-lzfbfCclutser-bound}),  we obtain the achievable normalized group spectral efficiency
given by (\ref{eq:bin-rate-zfbf-lb}).\footnote{A lower bound to an achievable rate is also achievable.}

\newpage

%\bibliographystyle{IEEEtran}
%\bibliography{biblio}
\bibliography{TDD-massive-v3.1}

\newpage

\end{document}